\documentclass[namedreferences]{SolarPhysics}
\usepackage[optionalrh,solaenum]{spr-sola-addons} 
\usepackage{graphicx}                    
\usepackage{amssymb}                    
\usepackage{mathrsfs}
\usepackage{color}                       
\usepackage{url}                         

\usepackage{txfonts}

\newcommand{\da}{\partial}
\newcommand{\alf}{Alfv\'{e}n }

\begin{document}

\begin{article}

\begin{opening}

\title{Resonant Absorption of Fast Magnetoacoustic Waves due
to Coupling into the Slow and Alfv\'{e}n Continua in the Solar Atmosphere}

\author{C.~T.~M.~\surname{Clack}$^{1}$\sep
        I.~\surname{Ballai}$^{1}$\sep
        M.~\surname{Douglas}$^{1}$}
\runningauthor{Clack et al.}
\runningtitle{Resonant absorption of FMA waves due to coupling into the slow and Alfv\'{e}n continua in the solar atmosphere}

\date{Received 18 February 2010; accepted 06 May 2010}

   \institute{$^{1}$Solar Physics and Space Plasma Research Centre (${\rm SP}^2{\rm RC}$), Department of Applied Mathematics,
University of Sheffield, Hounsfield Road, Hicks Building, Sheffield, S3 7RH, UK\\
    email: \url{app06ctc@sheffield.ac.uk}, \url{I.Ballai@sheffield.ac.uk}, \url{Mark.Douglas@sheffield.ac.uk}\\}

\begin{abstract}
Resonant absorption of fast magnetoacoustic (FMA) waves in an inhomogeneous, weakly dissipative, one-dimensional planar, strongly anisotropic and dispersive plasma is investigated. The magnetic configuration consists of an inhomogeneous magnetic slab sandwiched between two regions of semi-infinite homogeneous magnetic plasmas. Laterally driven FMA waves penetrate the inhomogeneous slab interacting with the localised slow or Alfv\'{e}n waves present in the inhomogeneous layer and are partly reflected, dissipated and transmitted by this region. The presented research aims to find the coefficient of wave energy absorption under solar chromospheric and coronal conditions. Numerical results are analyzed to find the coefficient of wave energy absorption at both the slow and Alfv\'{e}n resonance positions. The mathematical derivations are based on the two simplifying assumptions that (i) nonlinearity is weak, and (ii) the thickness of the inhomogeneous layer is small in comparison to the wavelength of the wave, \emph{i.e.} we employ the so-called long wavelength approximation. Slow resonance is found to be described by the nonlinear theory, while the dynamics at the \alf resonance can be described within the linear framework. We introduce a new concept of coupled resonances, which occurs when two different resonances are in close proximity to each other, causing the incoming wave to \emph{act as though} it has been influenced by the two resonances simultaneously. Our results show that the wave energy absorption is heavily dependent on the angle of the incident wave in combination with the inclination angle of the equilibrium magnetic field. In addition, it is found that FMA waves are very efficiently absorbed at the \alf resonance under coronal conditions. Under chromospheric conditions the FMA waves are far less efficiently absorbed, despite an increase in efficiency due to the coupled resonances.
\end{abstract}
\keywords{Magnetohydrodynamics (MHD); Sun: atmosphere; Sun: magnetic field; Sun: waves}
\end{opening}

\section{Introduction}

The complicated interaction of the motion of plasma with the magnetic fields is one of the most interesting processes in the solar atmosphere. A highly non-uniform and dynamical system, such as the solar atmosphere, is a perfect medium for magnetohydrodynamic (MHD) waves. These waves are able to transport momentum and energy along and across the magnetic field which can be dissipated, through a series of mechanisms, one of the most efficient being resonant absorption. The existence of resonant absorption is due to the coupling of global oscillations to local waves in a non-ideal plasma.

At resonance interacting dynamical systems can transfer energy to each other. In this context, resonant absorption has been suggested as a mechanism to create supplementary heating of fusion plasmas, but was, however, later rejected due to technical difficulties (see, {\emph e.g}. \opencite{tataronis1973}; \opencite{grossmann1973}; \opencite{Chen1974}; \opencite{Hasegawa1976}). \cite{ionson1978} suggested, for the first time, that resonant MHD waves may be a means to heat magnetic loops in the solar corona. Since then, resonant absorption of MHD waves has become a popular and successful mechanism for providing some of the heating of the solar corona (see, \emph{e.g.} \opencite{poedts1989}; \opencite{poedts1990a}; \opencite{poedts1990b}; \opencite{poedts1990c}; \opencite{ofman1995}; \opencite{belien1999}).

More recently resonant absorption has acquired a new applicability when the observed damping of waves and oscillations in coronal loops has been attributed to resonant absorption. Hence, resonant absorption has become a fundamental constituent block of one of the newest branches of solar physics, called \textit{coronal seismology} (which was first suggested by \opencite{uchida1973}; see also \opencite{roberts1984}). It has since been applied to explain the rapid damping of kink oscillations in coronal loops by numerous authors (see, \emph{e.g.} \opencite{nakariakov1999a}; \opencite{aschwanden1999}; \opencite{nakariakov2001}; \opencite{ruderman2002}; \opencite{goossens2002}; \opencite{terradas2008}; \opencite{terradas2010}). In addition, resonant absorption has been proposed as a mechanism to explain the observed loss of energy of acoustic oscillations in the vicinity of sunspots (see, \emph{e.g.} \opencite{hollweg1988}; \opencite{lou1990}; \opencite{sakurai1991}; \opencite{goossens1992A}; \opencite{erdelyi1995}; \opencite{goossens1995}).

The effect of resonant absorption has mostly been studied in linear MHD, mainly because the linear approach is easier to handle mathematically. However, linear theory shows that in the vicinity of the resonant position, the amplitude of oscillations can get large even when dissipative effects are taken into account resulting in a breakdown of the linear approach. The seminal study by \cite{ruderman1997c} showed that this is exactly what happens at the slow resonance in isotropic plasmas. The amplitude of oscillations far away from the resonance are small, but close to the resonant position nonlinearity must be taken into account. The study was then extended by \cite{ballai1998c} to anisotropic plasmas characteristic for the solar corona. More recently, \cite{clack2008} expanded the theory to take into account dispersion due to Hall currents.  All analytical studies on nonlinear resonant absorption focussed on resonant slow waves despite \alf waves being considered more important in the process of heating. Recently, \cite{clack2009d} showed that the \alf resonance can be described within linear MHD framework as the effects of largest nonlinear and dispersive terms identically cancel.

Observations have shown that the solar atmosphere is not only an inhomogeneous and structured medium, but is also highly dynamic. Solar magnetic structures, such as coronal loops, sunspots and prominences, exhibit large-scale motions (flows). Modern high resolution satellites, however, are still not accurate enough to derive the detailed diagnostics of these flows (present day observations are limited to a resolution of approximately 500 km, however, to fully resolve the processes creating these flows we could need a resolution of, conservatively, 1 km). On the other hand, theoretical studies have shown that these motions can considerably change the efficiency of MHD wave dissipation (see, \emph{e.g.} \opencite{erdelyi1996A}). The effect of steady (shear) flows on the resonant behaviour of nonlinear slow waves was first discussed by \cite{ballai1998a}.

Many studies of resonant absorption consider only the sound (or slow) and Alfv\'{e}n waves as excellent candidates for chromospheric and coronal heating (see, \emph{e.g.} \opencite{ruderman1997b}; \opencite{ballai1998b}; \opencite{ruderman2000}; \opencite{vasquez2005}). However, Alfv\'{e}n waves can only carry energy along the magnetic field lines and slow waves are only capable of carrying $1-2\%$ of their energy under coronal (low plasma-$\beta$) conditions. In contrast, fast magnetoacoustic (FMA) waves may have an important role in explaining the upper chromospheric and coronal temperatures as has been shown by, \emph{e.g.} \cite{cadez1997}; \cite{csik1998} and \cite{erdelyi2001}. FMA waves are magnetic waves which can propagate \emph{across} the magnetic field lines. They are compressional so they are subject to dissipation by, \emph{e.g.}, viscosity, heat conduction, Landau and transit-time damping, in the high-frequency limit where the Coulomb collisions are ineffective.

It is well known that in order to have acceptable heating by FMA waves coming from lower regions (\emph{e.g.} generated by convective motions below the photosphere) the waves should not be reflected by the steep rise of the Alfv\'{e}n and / or slow wave speed with height, nor should they become evanescent. However, this is difficult to prevent. On the other hand, to interact resonantly with the slow waves, FMA waves would only have to propagate to the upper chromosphere, since the ratio between characteristic speeds (slow and Alfv\'{e}n) is much smaller than in the solar corona. Only the high-frequency waves of the full FMA spectrum can reach the corona (with periods of a few tens of seconds). The energy flux density of fast waves at the bottom of the corona required for significant heating is consistent with the upper limit on acoustic waves (\opencite{athay1978}).

There is an alternative to the FMA waves propagating from the photosphere towards the solar corona. FMA waves (as well as slow and Alfv\'{e}n waves) may be generated locally in the chromosphere and corona by, \emph{e.g.} phenomena involving magnetic reconnection. It was suggested by \cite{parker1988} that shuffling the magnetic field in the solar atmosphere (by convective motion in lower regions) builds up magnetic stresses which can be released through, \emph{e.g.} reconnection providing the energy to maintain the high temperature of the solar upper atmosphere. It has been shown that reconnection events can produce MHD waves (see, \emph{e.g.} \opencite{roussev2001b,roussev2001c,roussev2001a}) through the process of \emph{nanoflaring}. The required amount of nanoflares has not yet been observed due to their localised nature. Even extending the theory to larger observable scales (explosive events, micro-flares, blinkers, etc.) still does not prove this theory (\opencite{aschwanden1999c}). Even though the energy released via reconnection may not be adequate to provide the heating required, it is still assumed that the amount of magnetic energy built up by the shuffling can provide enough energy to heat the upper atmosphere. The necessary supplementary heating may come from reconnection-driven, locally generated MHD waves.

The present paper has three major aims. First, is to study the nonlinear resonant interaction of FMA waves coupled into the slow continua in strongly anisotropic and dispersive plasmas (conditions typical of the solar upper chromosphere).
Secondly, to numerically investigate the resonant absorption of FMA waves at the Alfv\'{e}n resonance for conditions typically found in the solar corona. Finally, we numerically analyse the resonant absorption of FMA waves at a coupled (slow and Alfv\'{e}n) resonance in the solar chromosphere. The paper takes advantage of the analytical study by \cite{clack2009a} where they derived the coefficient of wave energy absorption for FMA, but did not discuss the implications of their findings in full based on concrete solar physics applications.

The paper is organized as follows. The next section describes the equilibrium model and any mathematical and / or physical assumptions made. In Section 3 we give the definition of the coefficient of wave energy absorption at the slow and Alfv\'{e}n resonant positions. Section 4 displays our numerical results and our interpretation of the derived coefficients. Finally, in Sect. 5 we summarize, draw our conclusions and discuss the applicability of the study.

\section{The Equilibrium and Assumptions}

\begin{figure}[!tb]
  \centering
  \includegraphics[width=7.5cm]{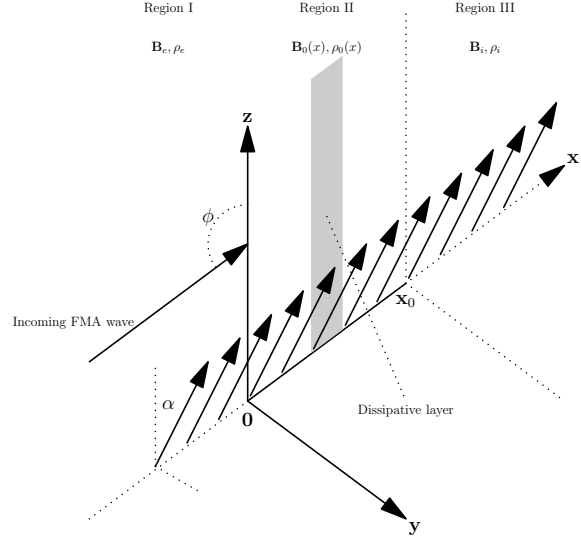}\\
  \caption{Schematic representation of the equilibrium state. Regions I ($x<0$) and III ($x>0$) contain a homogeneous magnetised plasma and Region II ($0<x<x_0$) an inhomogeneous magnetised plasma. The shaded strip shows the dissipative layer embracing the ideal resonant position $x_{\rm c(a)}$. Image reproduced from Clack and Ballai (2009b).}\label{fig:paperdiag1c}
\end{figure}

The magnetic field and density topology of the solar atmosphere is highly complex and, therefore, there are many degrees of freedom. However, to make our investigation tractable mathematically, we study the interaction of incident FMA waves in a one-dimensional plasma. The dynamics and absorption of waves will be studied in a Cartesian coordinate system. The configuration consists of an inhomogeneous and dissipative magnetised plasma $0<x<x_0$ (Region II) sandwiched between two semi-infinite regions with homogeneous and ideal magnetised plasmas for $x<0$ and $x>x_0$ (Regions I and III, respectively). The equilibrium state is shown in Figure \ref{fig:paperdiag1c}, which has been reproduced from \cite{clack2009a}. The equilibrium density and pressure are denoted by $\rho$ and $p$. The equilibrium magnetic field, $\mathbf{B}$, is unidirectional and lies in the $yz$-plane. In what follows the subscripts ``e", ``0" and ``i" denote the equilibrium quantities in the three regions (Regions I, II, III, respectively). It is convenient to introduce the inclination angle, $\alpha$, between the $z$-axis and the direction of the equilibrium magnetic field, so that the components of the equilibrium magnetic field are
\begin{equation}\label{eq:magfieldequilibrium}
B_y=B\sin\alpha,\mbox{ }B_z=B\cos\alpha.
\end{equation}
All equilibrium quantities are continuous at the boundaries of Region II, and they satisfy the equation of total pressure balance. It follows from the equation of total pressure balance that the density ratio between Regions I and III satisfy the relation
\begin{equation}\label{eq:densityrat}
\frac{\rho_{\rm i}}{\rho_{\rm e}}=\frac{2c_{\rm Se}^2+\gamma
v_{\rm Ae}^2}{2c_{\rm Si}^2+\gamma v_{\rm Ai}^2},
\end{equation}
where the Alfv\'{e}n and sound speeds are
\begin{equation}\nonumber
v_{\rm A}=\frac{B}{\sqrt{\mu_0\rho}}\mbox{ }
{\rm and}\mbox{ }c_{\rm S}=\sqrt{\frac{\gamma p}{\rho}},
\end{equation}
with $\gamma$ being the adiabatic exponent and $\mu_0$ is the permeability of free space.

We consider an equilibrium such that the plasma in Region III is both hotter and more rarefied than in Region I and assume a simple monotonic linear profile for all variables in Region II. This choice provides a single unique resonant surface for both the slow and \alf resonances. The objective of the present paper is to study the resonant absorption of FMA waves at slow and Alfv\'{e}n dissipative layers. There are two cases which we wish to investigate. First, we study the absorption of fast waves at Alfv\'{e}n resonance as a possible scenario of the interaction of global fast waves (modelling EIT waves) and coronal loops. Secondly, we analyse the absorption of fast waves at the slow resonance as a possible scenario for heating in the upper chromosphere. When a fast wave interacts with a slow resonance, an Alfv\'{e}n resonance is also present. We consider this scenario in our analysis and study the total (\emph{i.e.} coupled) absorption in the inhomogeneous layer.

In an attempt to remove other effects from the analysis we consider the incoming fast wave to be entirely in the $xz$-plane, \emph{i.e.} $k_y=0$. The resonant positions are located where the global fast wave speed coincides with that of the local slow / Alfv\'{e}n speed. The dispersion relation for the impinging propagating fast waves takes the form
\begin{equation}\label{eq:dispersion1}
\frac{\omega^2}{k^2}=\frac{1}{2}\left\{\left(v_{\rm A}^2+c_{\rm S}^2\right)
+\left[\left(v_{\rm A}^2+c_{\rm S}^2\right)^2-4v_{\rm A}^2c_{\rm S}^2\cos^2\phi\right]^{1/2}\right\},
\end{equation}
where $\phi$ is the angle between the direction of propagation and the background magnetic field within the $xz$-plane and
$k=(k_x^2+k_z^2)^{1/2}$. For the sake of simplicity, the ratio $k_x/k_z$ is denoted $\kappa_{\rm e}$ so that the dispersion relation (\ref{eq:dispersion1}) becomes
\begin{equation}\label{eq:dispersion2}
\frac{\omega^2}{k^2}=\frac{1}{2}
\left\{\left(v_{\rm A}^2+c_{\rm S}^2\right)+\left[\left(v_{\rm A}^2+c_{\rm S}^2\right)^2-4\frac{v_{\rm A}^2c_{\rm S}^2}{1+\kappa_{\rm e}^2}\right]^{1/2}\right\},
\end{equation}
where $1+\kappa_{\rm e}^2=1/\cos^2\phi$.

We assume the plasma is \emph{strongly} magnetized in the three regions, such that the conditions $\omega_{\rm i(e)}\tau_{\rm i(e)}\gg1$ are satisfied, where $\omega_{\rm i(e)}$ is the ion (electron) gyrofrequency and $\tau_{\rm i(e)}$ is the ion (electron) collision time. Due to the strong magnetic field, transport processes are derived from Braginskii's stress tensor (see, \emph{e.g.} \opencite{braginskii1965}; \opencite{erdelyi1995}; \opencite{ruderman1997i}; \opencite{ruderman1997a}; \opencite{mocanu2008}). As we deal with two separate waves (slow and Alfv\'{e}n), we will need to choose the particular dissipative process which is most efficient for these waves. For slow waves, it is a good approximation to retain only the first term of Braginksii's expression for viscosity, namely compressional viscosity (\opencite{Hollweg1985}). In addition, in the solar upper atmosphere slow waves are sensitive to thermal conduction. In a strongly magnetized plasma, the thermal conductivity parallel to the magnetic field lines dwarfs the perpendicular component, hence the heat flux can be approximated by the parallel component only (\opencite{priest1984}). On the other hand, since Alfv\'{e}n waves are transversal and incompressible they are affected by the second and third components of Braginskii's stress tensor, called shear viscosity (\opencite{clack2009d}). Finally, Alfv\'{e}n waves are efficiently damped by finite electrical conductivity, which becomes anisotropic under coronal conditions. The parallel and perpendicular components, however, only differ by a factor of 2, so we will only consider one of them without loss of generality. All other transport mechanisms can be neglected.

The dynamics of nonlinear resonant MHD waves in anisotropic and dispersive plasmas was studied by \cite{clack2008} and \cite{clack2009d}. They derived the governing equations and connection formulae necessary to study resonant absorption in slow / Alfv\'{e}n dissipative layers. A further paper by \cite{clack2009a} studied the interaction of fast waves with the slow and Alfv\'{e}n resonances and derived the coefficients of wave energy absorption. They intentionally overlooked the complication of an Alfv\'{e}n resonance being present at the same time as the slow resonance. In the present paper, however, we include both resonances. These conditions would be typical of the upper chromosphere where the plasma-$\beta$ goes from being larger than unity to becoming smaller than unity, retaining the monotonic dependence of characteristic speeds in the inhomogeneous layer.

The key assumptions we use in the present paper in order to facilitate analytical progress are connected to the strength of nonlinearity and the wavelength of the incoming wave. First, from the very beginning we assume that nonlinearity, at the slow resonance, is weak. This simplification allows us to study the nonlinear governing equation derived by \cite{clack2008} analytically. \cite{ruderman2000} investigated the absorption of sound waves at the isotropic slow dissipative layer in the limit of strong nonlinearity. In their analysis nonlinearity dominated dissipation in the \emph{resonant layer} which embraces the dissipative layer. They concluded that nonlinearity decreases absorption in the long wavelength approximation, but increases it at intermediate values of $kx_0$, however, the increase is never more than $20\%$. Secondly, we also assume that the thickness of the inhomogeneous region (Region II) is thin in comparison to the wavelength of the impinging fast wave, \emph{i.e.} $kx_0\ll1$, which has two implications. The first is to enable us to neglect terms of the order of $k^2x^2_0$ (and above) in the calculations. The second consequence is more subtle. To have a coupled resonance, the incoming FMA wave must resonantly interact with an \alf resonance and be partially transmitted and then resonantly interact with a slow resonance. On one hand, if the \alf and slow resonances are more than one wavelength, $k^{-1}$, apart the transmitted wave will have decayed by the time it reaches the slow resonance. On the other hand, if the \alf and slow resonance are within one wavelength of each other then the transmitted wave will be able to interact with the slow resonance. So, if an environment is set up such that both an \alf and a slow resonance occur within the inhomogeneous region the distance between the resonances, $x_1$, must satisfy the condition $x_1<x_0$. Hence, we have $kx_0\ll1\Rightarrow x_0\ll k^{-1}\Rightarrow x_1\ll k^{-1}$.

We will use the same governing equations, absorption coefficients and results as obtained by \cite{clack2009a}. We will only derive new results necessary to investigate the processes at the coupled resonance.

\section{Coefficient of Wave Energy Absorption}

The basic premise of our model is that one single monochromatic wave (the FMA wave) interacts with the inhomogeneous layer and one (or more) wave leaves it. The difference of the energy flux entering and leaving the resonance quantifies the energy resonantly absorbed inside the inhomogeneous region. This energy can then be converted into heat by dissipation. The coefficient of wave energy absorption is just the \emph{percentage} of the energy entering the inhomogeneous region that is resonantly absorbed.

\cite{clack2009a} derived the coefficient of wave energy absorption of fast magnetoacoustic waves at both the Alfv\'{e}n and slow resonances, but the derivation requires a long and cumbersome mathematical treatment so we do not repeat their working, we just recall their main results. It should be noted that when they derived the coefficients of wave energy absorption they relied on connection formulae, which are assumed to be accurate (\opencite{stenuit1995}). At the Alfv\'{e}n resonance the wave dynamics are described by linear theory, as proved by \cite{clack2009d}. At resonance, the incoming monochromatic FMA wave is partly reflected, transmitted and dissipated. The reflected wave is monochromatic and the transmitted wave carries no energy because it is evanescent in Region III. The coefficient of wave energy absorption can be written as (see, \emph{e.g.} \opencite{ruderman1997b}; \opencite{clack2009a})
\begin{equation}\label{eq:Gamma}
\Gamma=\frac{\Pi_{\rm{in}}-\Pi_{\rm{out}}}{\Pi_{\rm{in}}}=1-\frac{1}{p_{\rm e}^2}\sum_{n=1}^{\infty}|\overline{A}_n|^2,
\end{equation}
where $\Pi_{\rm{in}}$ and $\Pi_{\rm{out}}$ are the normal components of the energy fluxes, averaged over a period,
of the incoming and outgoing waves, respectively. Here $p_{\rm e}$ is the plasma pressure in Region I and $\overline{A}_n$ denotes the \emph{total} amplitude of the reflected wave at that order of approximation. Thus, the coefficient of wave energy absorption of FMA waves at the Alfv\'{e}n resonance is given by
\begin{equation}
\Gamma_{\rm{a}}=\frac{4\tau_{\rm a}\mu}{(\tau_{\rm a}+\mu)^2+\upsilon^2}+\mathscr{O}(k^2x_0^2),
\end{equation}
where $\tau_{\rm a}$, $\mu$ and $\upsilon$ represent
\begin{equation}\label{eq:taumuupsilon}
\tau_{\rm a} =\frac{\pi kV\sin^2\alpha}{\rho_{0{\rm a}}|\Delta_{\rm a}|},\quad
\mu =\frac{\kappa_{\rm e}V}{\rho_{\rm e}\left(V^2-v_{\rm Ae}^2\cos^2\alpha\right)},
\end{equation}
\begin{equation}\label{eq:upsilon111}
\upsilon =\frac{\kappa_{\rm i}V}{\rho_{\rm i}\left(V^2-v_{\rm Ai}^2\cos^2\alpha\right)}
-kV\mathscr{P}\int_{0}^{x_0}F^{-1}(x)\mbox{ }dx.
\end{equation}
Here the subscript `a' means an equilibrium quantity that has been calculated at the \alf resonant surface, $V$ is the phase velocity of the incoming FMA wave, given by Equation (\ref{eq:dispersion1}), $\kappa_{\rm e}=\pm\tan\phi$, $\Delta_{\rm a}=-(dv^2_{\rm A}/dx)_{\rm a}\cos^2\alpha$ and $\kappa_{\rm i}$ is defined as
\begin{equation}\nonumber
\kappa_{\rm i}=\sqrt{-\frac{V^4-V^2\left(c_{\rm Si}^2+v_{\rm Ai}^2\right)+c_{\rm Si}^2v_{\rm Ai}^2\cos^2\alpha}
{\left(c_{\rm Si}^2+v_{\rm Ai}^2\right)\left(V^2-c_{\rm Ti}^2\cos^2\alpha\right)}}.
\end{equation}
The Cauchy principal part, $\mathscr{P}$, is used in the definition of $\upsilon$ because the integral is divergent. The function $F(x)$ is the coefficient function of the highest order derivative in the Hain--Lust equation defining the continuous parts in the linear spectrum. It takes the form
\begin{equation}\nonumber
F(x)=\frac{\rho_0\left(v_{\rm A}^2+c_{\rm S}^2\right)\left(V^2-v_{\rm A}^2\cos^{2}\alpha\right)\left(V^2-c_{\rm T}^2\cos^{2}\alpha\right)}{V^4-V^2\left(v_{\rm A}^2+c_{\rm S}^2\right)+v_{\rm A}^2c_{\rm S}^2\cos^{2}\alpha},
\end{equation}
with $c_{\rm T}=c_{\rm S}v_{\rm A}(c_{\rm S}^2+v_{\rm A}^2)^{-1/2}$ being the cusp (tube) speed.

The governing equation at the slow resonance have has been derived by \cite{clack2008}. Similar to the Alfv\'{e}n resonance the incoming FMA wave is partly reflected, transmitted and dissipated. The transmitted wave is decaying in a similar way as before, however, the reflected wave includes higher harmonics in addition to the fundamental mode due to the nonlinearity at the resonance. Using the approximation of weak nonlinearity, in the first order of approximation, we find the fundamental mode of the reflected wave. In the second order approximation, we see that the higher harmonics are of the order of $k^2x_0^2$. In the third order of approximation, we find an additional contribution to the reflected fundamental mode due to nonlinearity and dispersion. For even higher orders of approximation the contributions are no longer monochromatic, therefore, they are neglected (for full details on these estimations see, \emph{e.g.} \opencite{clack2009a}). The coefficient of wave energy absorption of FMA waves at the slow resonance is
\begin{equation}\label{slowabs}
\Gamma_{\rm{c}}=\frac{4\tau_{\rm c}\mu}{\mu^2+\upsilon^2}-\zeta^2
\frac{27p_{\rm e}^2\tau_{\rm c}^3\mu^3\cos^4\alpha}{6\pi^2V^2k^2x_0^2\left(\mu^2+\upsilon^2\right)^2}
+\mathscr{O}(k^2x_0^2),
\end{equation}
where $\mu$ and $\upsilon$ take the same form as in Equations (\ref{eq:taumuupsilon}) and (\ref{eq:upsilon111}), and $\tau_{\rm c}$ is derived as
\begin{equation}\label{eq:tauc}
\tau_{\rm c}=\frac{\pi kV^5}{\rho_{0{\rm c}}v_{\rm Ac}^4|\Delta_{\rm c}|\cos^2\alpha}.
\end{equation}
Here the subscript `c' means an equilibrium quantity has been calculated at the slow resonant surface and $\Delta_{\rm c}=-(dc^2_{\rm T}/dx)_a\cos^2\alpha$. The first term in Equation (\ref{slowabs}) is the linear coefficient of wave energy absorption, and the second term is the nonlinear correction (both terms are of the order of $kx_0$). The physical meaning of the small term $\zeta$ is buried within the nonlinear governing equation, but essentially describes the ratio of nonlinear and dissipative terms (for further details we refer to \opencite{clack2009a}). The small quantity $\zeta$ ensures that the contribution of nonlinearity to the net coefficient of absorption is small. In addition, since the nonlinear correction is always positive, it is clear that the effect of nonlinearity and dispersion is to decrease the coefficient of wave energy absorption, a result already obtained earlier by \cite{ballai1998b}. \cite{clack2009a} went on to show that when Hall dispersion is also taken into account, the nonlinear term becomes 26 times larger, further reducing the overall coefficient.

As mentioned earlier, the above discussed coefficients of wave energy absorption (taken separately) do not give the complete picture. In principle, every time a FMA wave interacts with a slow resonance an Alfv\'{e}n resonance is also present (it is easy to show that under coronal conditions, the vise-versa statement is not true). Usually, to avoid this happening we would align the magnetic field with the $z-$axis, and since $\da/\da y=0$ the Alfv\'{e}n resonance would vanish. However, to include Hall currents (which provides dispersive effects at the slow resonance) we must have an angle between the $z-$axis and the equilibrium magnetic field. In an attempt to tackle this problem, we investigate the interaction of FMA waves with an inhomogeneous region that contains \emph{both} a slow and an Alfv\'{e}n resonance. To carry this out analytically, we must assume that the thickness of the inhomogeneous region is much smaller than the wavelength of the incoming wave, \emph{i.e.} $kx_0\ll 1$.  We also assume that there is a single and unique slow and \alf resonance (this assumption can be relaxed later).

To find the coefficient of wave energy absorption at a coupled resonance the procedure is identical to that carried out in \cite{clack2009a}, so we write out our findings rather than presenting all calculations in detail. First, we found that, at a coupled resonance, in the first order approximation (\emph{i.e.} deriving the fundamental mode form), we cannot decouple the Alfv\'{e}n and slow reflected waves. This is in agreement with \cite{woodward1994a} and \cite{woodward1994b} who state that when Hall currents are present the slow and Alfv\'{e}n modes cannot always be decoupled. Even though these studies were investigating stationary waves, we believe the same applies here with resonant absorption, specifically, if we approximate that the wave interacts with both resonances simultaneously. The amplitude of the reflected wave in the first order of approximation is derived as
\begin{equation}\label{eq:use1}
A_a+A_{{\rm c}1}=-p_{\rm e}\frac{\tau-\mu+i\upsilon}{\tau+\mu+i\upsilon},
\end{equation}
where $\tau=\tau_{\rm a}+\tau_{\rm c}$, and $\mu$ and $\upsilon$ are given by Equations (\ref{eq:taumuupsilon}) and (\ref{eq:upsilon111}), respectively.
In higher order approximations we find that the Alfv\'{e}n resonance has no contribution to the amplitudes of the reflected waves. In particular, the addition to the fundamental mode due to nonlinearity and dispersion is identical to that found when no Alfv\'{e}n resonance is present, which is to be expected, since the dynamics at the Alfv\'{e}n resonance is linear and, therefore, should have no contribution to nonlinear effects.

Using Equation (\ref{eq:Gamma}), the coefficient of wave energy absorption of FMA waves inside the inhomogeneous region is calculated to be (in the long wavelength approximation)
\begin{equation}\label{eq:coeffmajor}
\Gamma=\frac{4\tau\mu}{(\tau+\mu)^2+\upsilon^2}
-\zeta^2\frac{27p_{\rm e}^2\tau_c^3\mu^3\cos^4\alpha}{6\pi^2V^2k^2x_0^2\left(\mu^2+\upsilon^2\right)^2}
+\mathscr{O}(k^2x_0^2).
\end{equation}

In the next section, we will numerically analyse the coefficient of wave energy absorption and, in particular, we investigate what is the significance of the angle of the incident wave and the inclination of the ambient magnetic field on the absorption of wave energy at the coupled resonance.

\section{Numerical Results}

Before proceeding to the mentioned analysis of the wave energy absorption, it is instructive to consider the contributions to $\tau$ due to the slow and Alfv\'{e}n resonances. We can easily calculate what percentage of $\tau$ comes from each resonance by studying Equation (\ref{eq:coeffmajor}). Immediately it is clear that the nonlinear term comes wholly from the slow resonance, as we have already discussed. The linear term is a combination of the resonant absorption due to the slow and \alf resonance. The percentage contributions of each resonance is found by calculating $\tau_{\rm c}/\tau$ ($\%$ of slow contribution) and $\tau_{\rm a}/\tau$ ($\%$ of \alf contribution). Using Equations (\ref{eq:taumuupsilon}), (\ref{eq:tauc}) and (\ref{eq:coeffmajor}) we obtain
\begin{equation}\label{eq:percentages}
\frac{\tau_{\rm c}}{\tau}=\frac{K_{1}}{K_{1}+K_{2}},\quad\frac{\tau_{\rm a}}{\tau}=\frac{K_{2}}{K_{1}+K_{2}},
\end{equation}
where
\begin{equation}\label{eq:K1K2}
K_{1}=\rho_{0{\rm a}}|\Delta_{\rm a}|V^4,\quad K_{2}=\rho_{0{\rm c}}v^{4}_{\rm Ac}|\Delta_{\rm c}|\sin^2\alpha\cos^2\alpha.
\end{equation}

\begin{figure}[!htb]
  \centering
  \includegraphics[width=5.5cm]{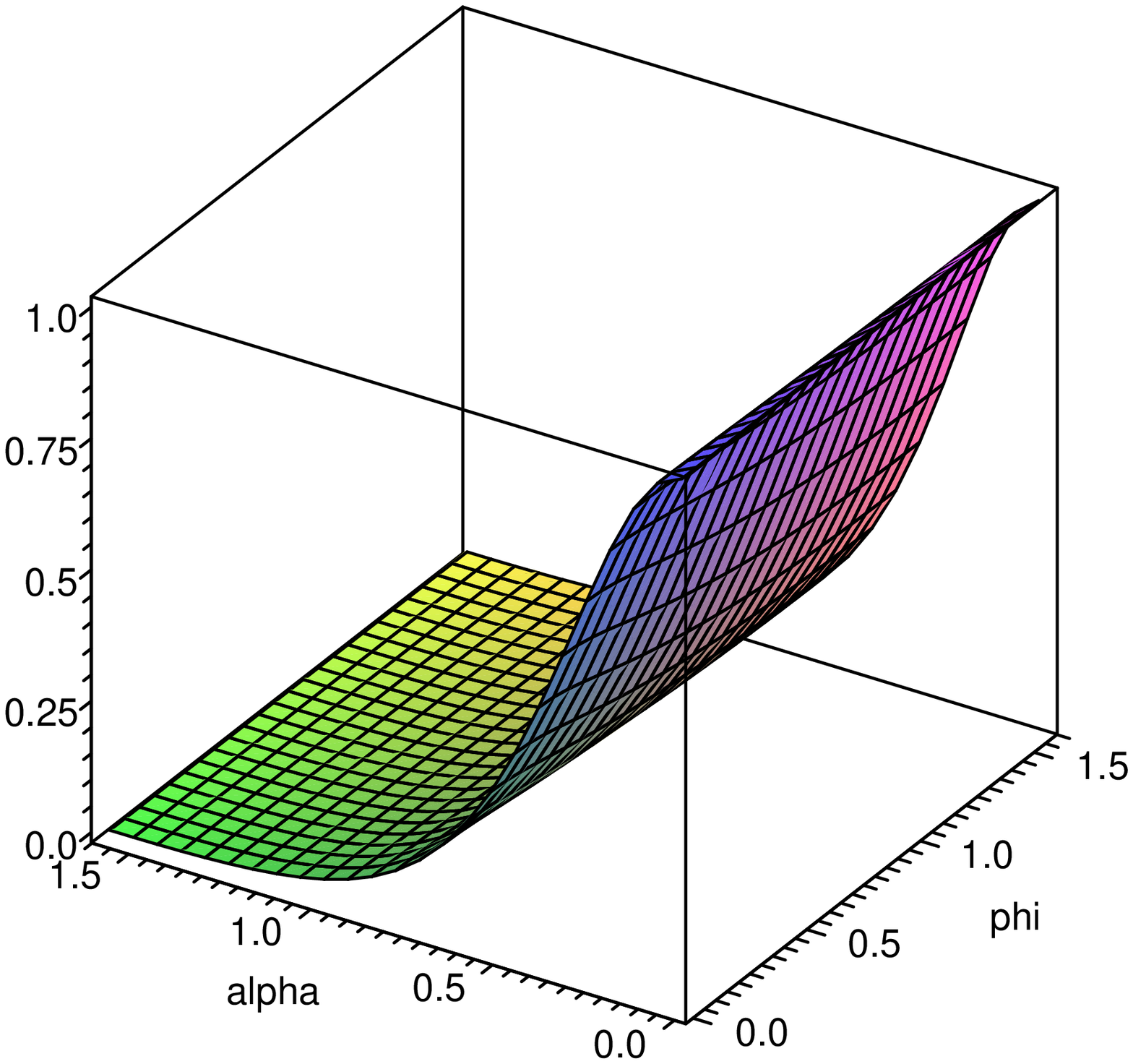}\hspace{2ex}\includegraphics[width=5.5cm]{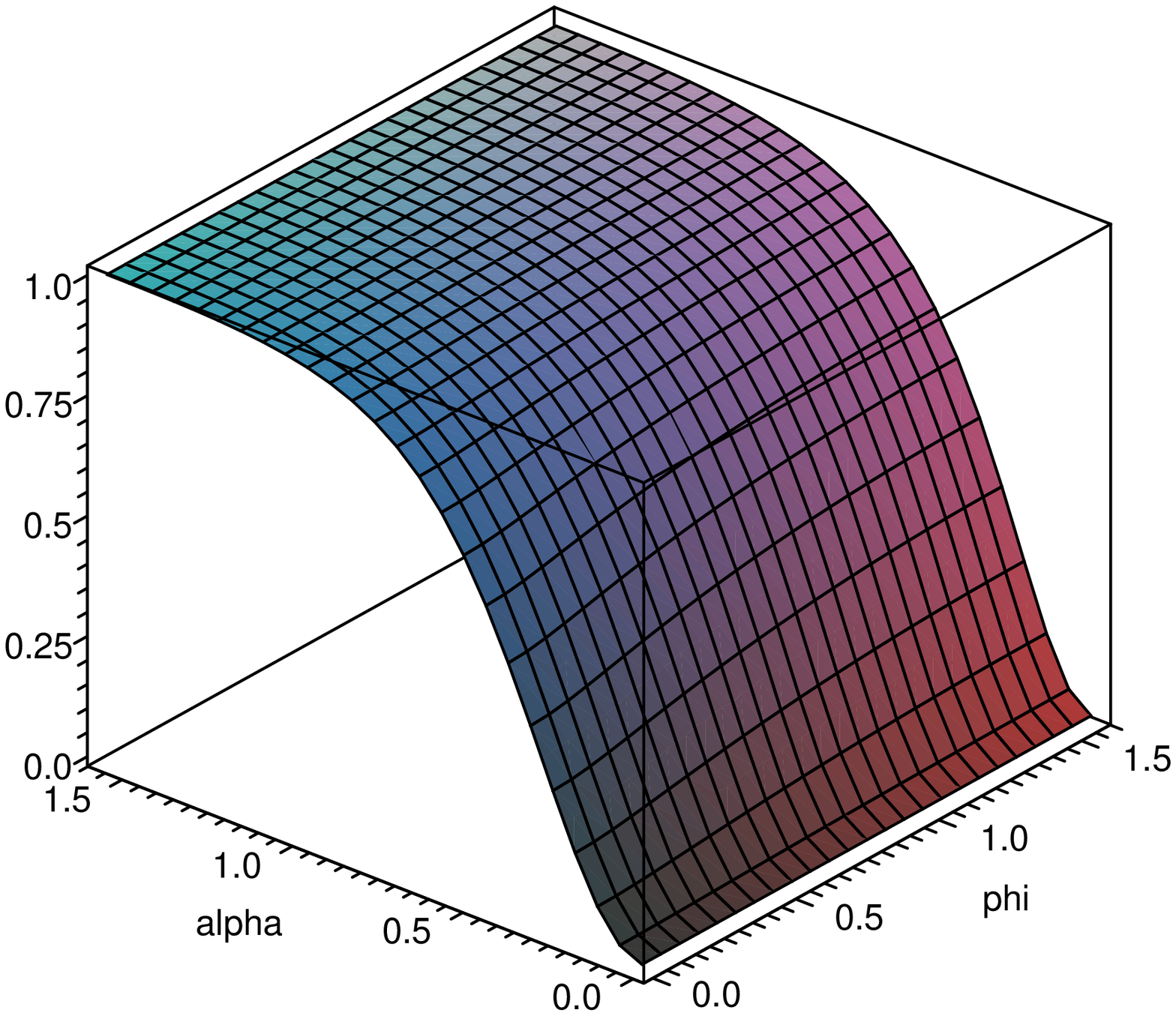}\\
  \caption{Left plot: percentage contribution to $\tau$ in the coupled inhomogeneous layer due to the slow resonance in terms of the wave incident angle, $\phi$, and the inclination angle of the magnetic field, $\alpha$. Right plot: percentage contribution to $\tau$ due to the \alf resonance in terms $\phi$ and $\alpha$.} \label{fig:slowcontrib}
\end{figure}

The particular form of Equation (\ref{eq:percentages}) depends entirely on the choice of profile for the equilibrium quantities inside the inhomogeneous layer. For our analysis, in Region II (the inhomogeneous region) we have chosen a monotonically increasing linear profile for all equilibrium quantities (including characteristic speeds) of the form $f(x)=f(0)+x/x_0[f(x_0)-f(0)]$. Using the values $v_{\rm Ae}=28\mbox{ }{\rm km\mbox{ }s}^{-1}$, $c_{\rm Se}=34\mbox{ }{\rm km\mbox{ }s}^{-1}$, $v_{\rm Ai}=156\mbox{ }{\rm km\mbox{ }s}^{-1}$, $c_{\rm Si}=65\mbox{ }{\rm km\mbox{ }s}^{-1}$  and $\rho_{\rm e}=5\times10^{-11}\mbox{ }{\rm kg\mbox{ }m}^{-3}$ for characteristic speeds we can obtain the variations given by Figure \ref{fig:slowcontrib}.
The left plot in Figure \ref{fig:slowcontrib} shows the percentage contribution to the resonant absorption due to the slow resonance, while the right plot shows the same for the \alf resonance. It is easily seen that $\phi$ does not have much effect on the contributions, however, $\alpha$ has a marked effect. At $\alpha=0$, there is no contribution from the \alf resonance - which is to be expected since, for this particular inclination, an \alf resonance does not exist. As the angle $\alpha$ increases the contribution due to the \alf resonance increases rapidly (and accordingly the contribution from the slow resonance decreases), and by $\alpha=\pi/4$ the \alf resonance contributes $80\%$ of the absorbed energy. At $\alpha=\pi/2$ the slow resonance disappears, and so its contribution drops to zero.

\subsection{Alfv\'{e}n Resonance: Modelling the Interaction of EIT Waves with Coronal Loops}

Fast waves that are generated by the convection motion and propagate upwards are reflected by the strong density gradients in the upper chromosphere, so just a tiny proportion of FMA waves are able to reach the corona. Fast waves, however, can be generated in the corona by, \emph{e.g.} flaring processes, in particular coronal mass ejections (CMEs). Global disturbances generated by CMEs, known as EIT (Extreme ultraviolet Imaging Telescope) waves, are believed to be FMA waves (\opencite{ballai2005}) propagating in the quiet Sun. In their propagation, EIT waves interact with active region loops setting them into motion (see, \emph{e.g.} \opencite{ballai2008}). The present section is devoted to the study of resonant coupling of FMA waves (modelling EIT waves) with local \alf waves in coronal loops. One major problem identified with the model we would like to use is that the EIT waves have very low speeds ($300-500\mbox{ }{\rm km\mbox{ }s}^{-1}$) compared with the accepted \alf speed outside coronal loops ($\geq1000\mbox{ }{\rm km\mbox{ }s}^{-1}$). The problem associated with the speed of EIT waves could be resolved by assuming a steady rise in temperature and magnetic field strength as an active region is approached. The rise in temperature increases the sound speed, while the increase in strength of magnetic field will cause the \alf speed to grow. Hence, the FMA wave speed will increase as it is a combination of the sound and \alf speeds. As the gradients are not steep enough to cause shocks or reflection, the EIT waves are accelerated to the local speeds as they approach active regions.

Another assumption our model relies on is that $v_{\rm Ai}>v_{\rm Ae}$, \emph{i.e.} the \alf speed in Region I is less than that in Region III. Since we can only integrate $F(x)$ when we assume monotonic functions inside the inhomogeneous region we have to chose whether these functions will increase or decrease. If the functions decrease in the inhomogeneous region there will be no absorption since the FMA waves will never resonantly interact with the local \alf waves. However, if the functions increase, we do achieve resonant interactions of the two modes. A more accurate model of reality at a coronal loop's edge would be a complicated non-monotonic function that both increases and decreases the local equilibrium parameters which could create resonant interactions. Essentially, there would be several resonant positions inside the inhomogeneous region, however, we cannot analytically solve such a model at present (due to the form that $F(x)$ would take). Therefore, to provide a valuable insight into the efficiency of resonant absorption at coronal loops, we choose an unrealistic but tractable model. Mathematically, the above restriction can be easily represented by the condition that $\max[v_{\rm A}(x)]>v_{\rm A_e}$ where $x\in (0,x_0]$. In addition, for a single unique resonance we need to impose the constraint $dv_{\rm A}(x)/dx\neq 0$, $x\in(0,x_0)$. It is also clear, from the outset, that under coronal conditions ($\beta\ll1$) FMA waves will never resonantly interact with local slow waves.

\begin{figure*}[!htb]
  \centering
  \includegraphics[width=5.5cm]{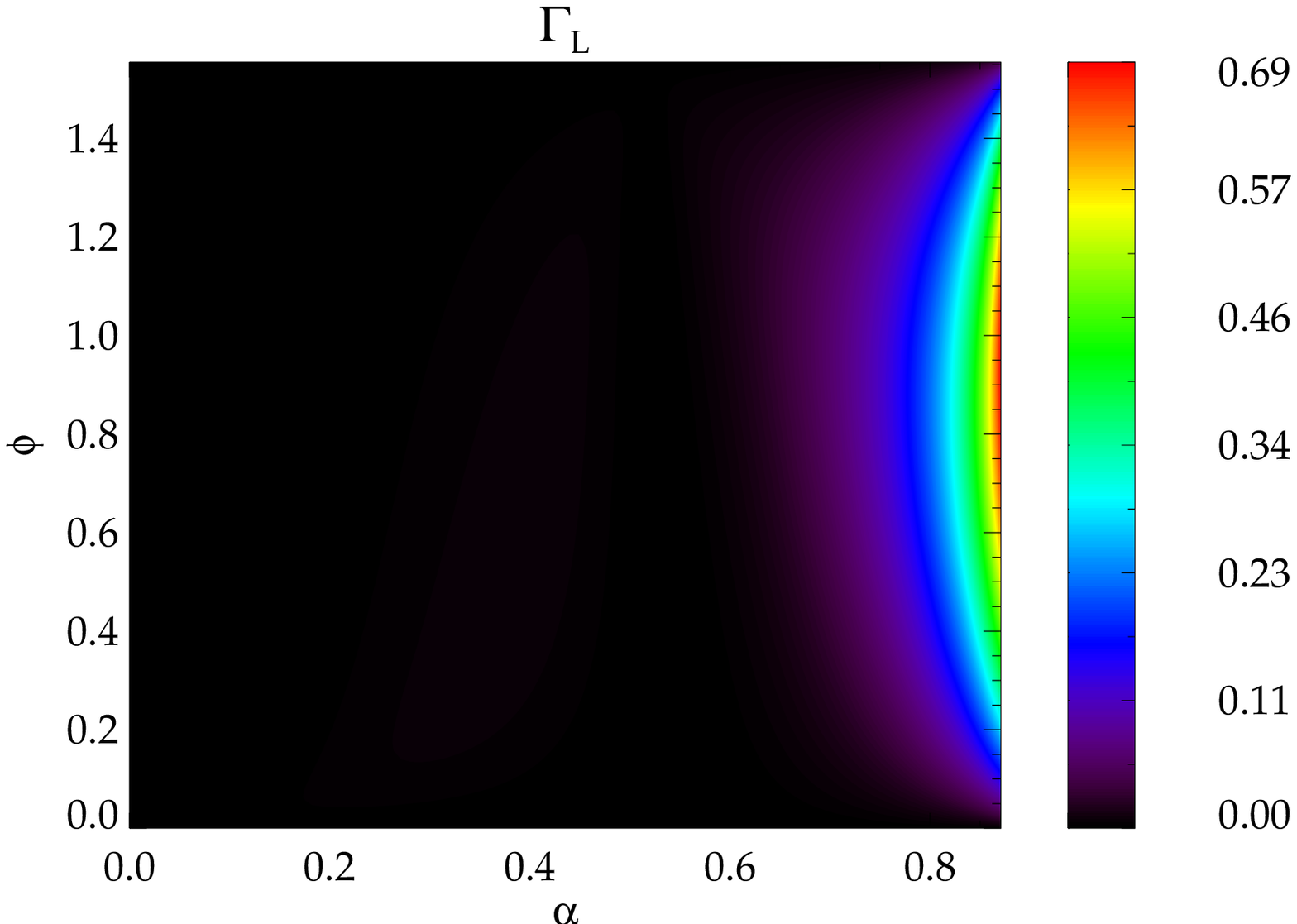}\hspace{2ex}\includegraphics[width=5.5cm]{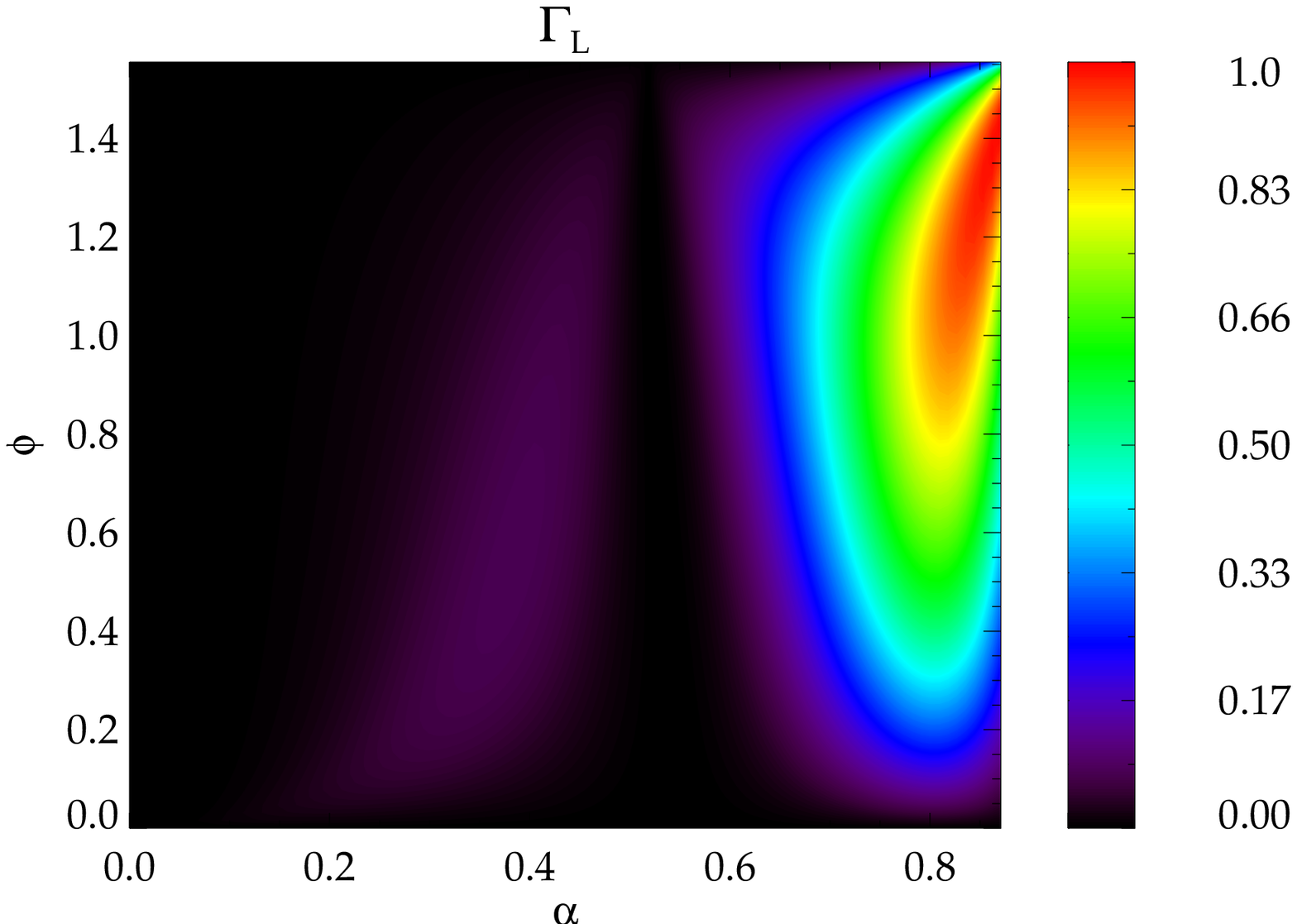}\\
  \includegraphics[width=5.5cm]{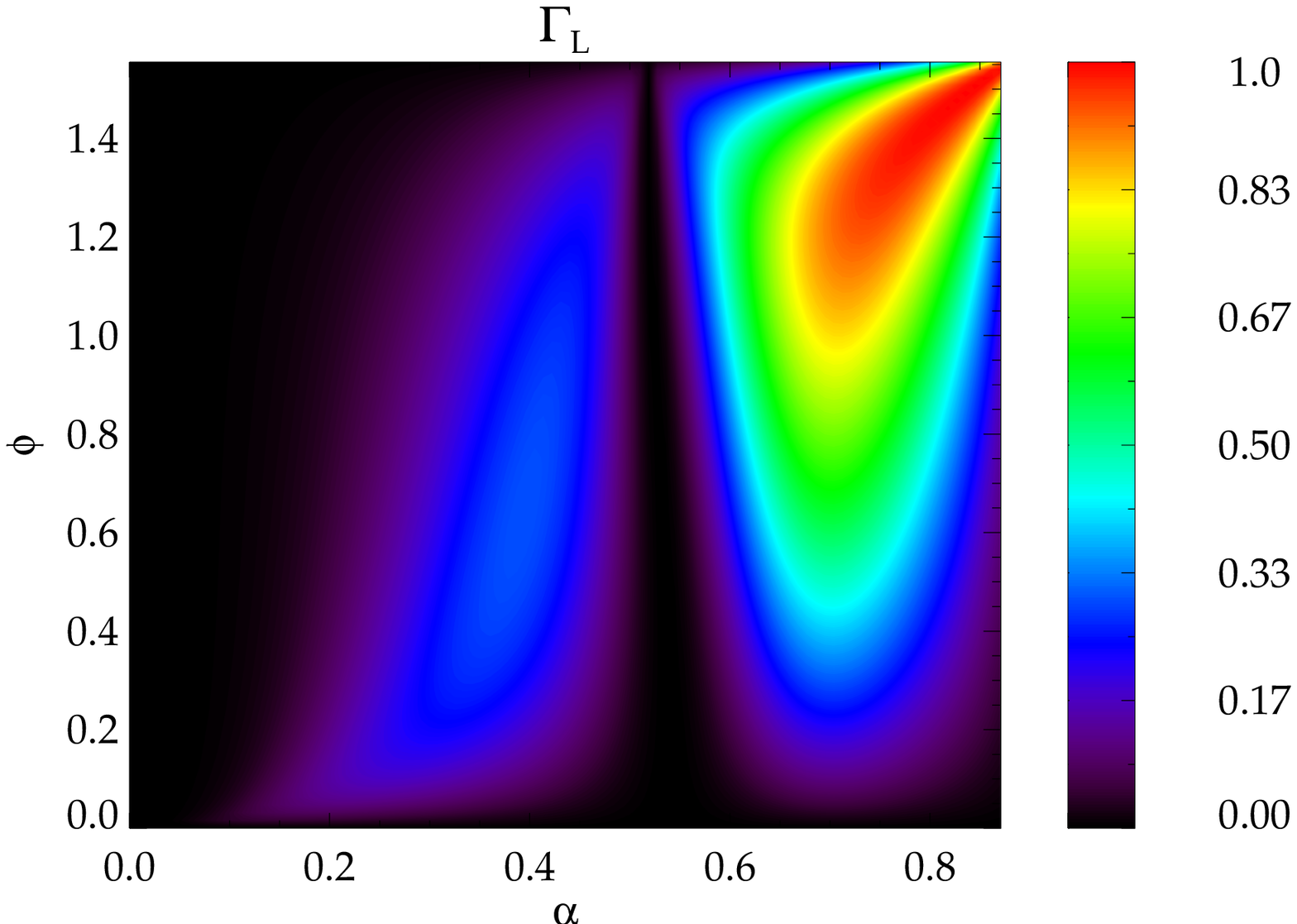}\hspace{2ex}\includegraphics[width=5.5cm]{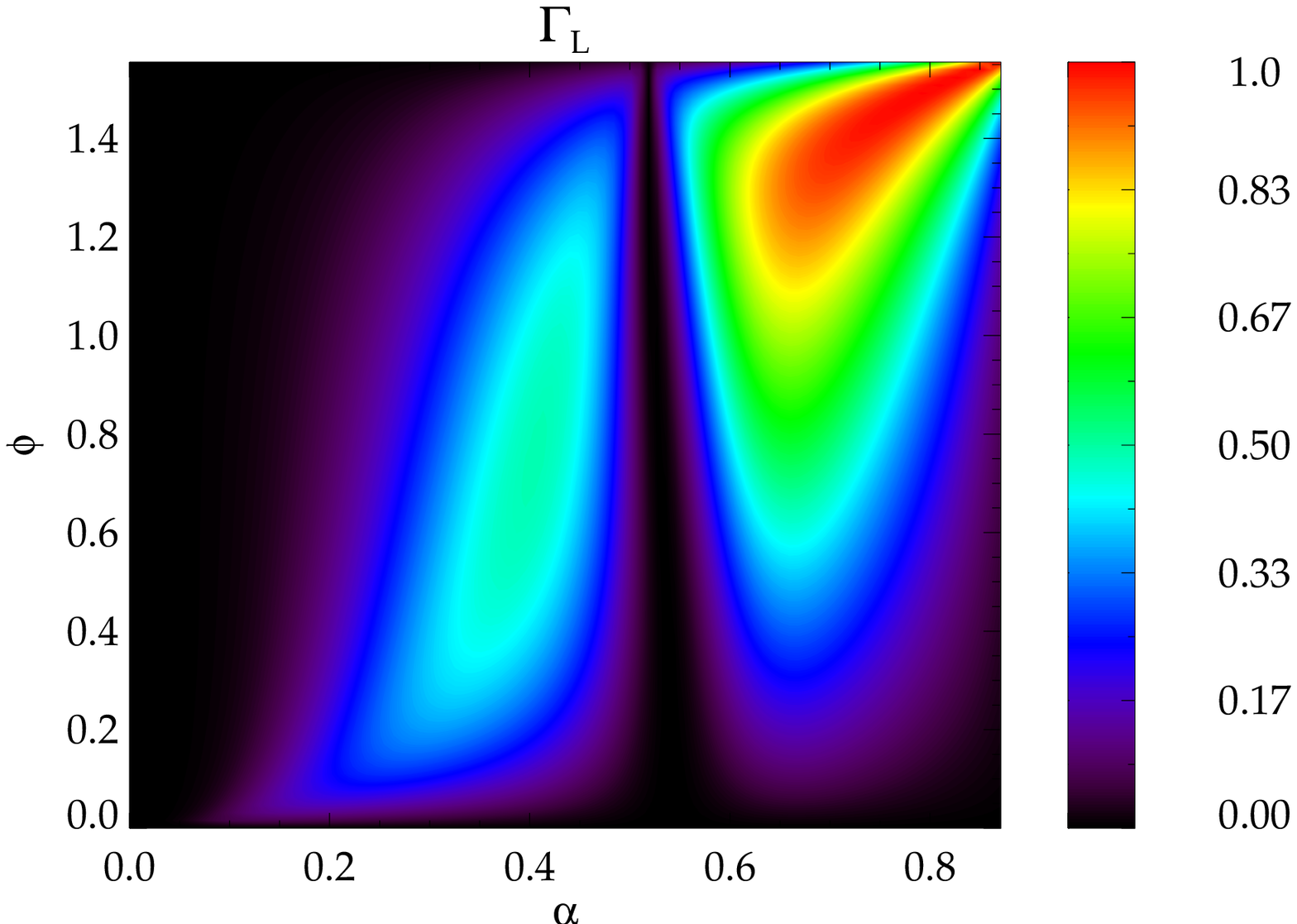}\\
  \caption{The wave energy absorption coefficient of FMA waves at the Alfv\'{e}n resonance. Here we have $v_{\rm Ae}=1200\mbox{ }{\rm kms}^{-1}$, $c_{\rm Se}=200\mbox{ }{\rm km\mbox{ }s}^{-1}$, $v_{\rm Ai}=1400\mbox{ }{\rm km\mbox{ }s}^{-1}$, $c_{\rm Si}=250\mbox{ }{\rm km\mbox{ }s}^{-1}$ and $\rho_{\rm e}=1.33\times10^{-12}\mbox{ }{\rm kg\mbox{ }m}^{-3}$. The dimensionless quantity, $kx_0$, takes the values (from top left to bottom right) $0.01$, $0.1$, $0.5$ and $1.0$, respectively.} \label{fig:linearalfven}
\end{figure*}

To simulate conditions typical of the solar corona we consider $v_{\rm Ae}=1200\mbox{ }{\rm km\mbox{ }s}^{-1}$, $c_{\rm Se}=200\mbox{ }{\rm km\mbox{ }s}^{-1}$, $v_{\rm Ai}=1400\mbox{ }{\rm km\mbox{ }s}^{-1}$, $c_{\rm Si}=250\mbox{ }{\rm km\mbox{ }s}^{-1}$ and $\rho_{\rm e}=1.33\times10^{-12}\mbox{ }{\rm kg\mbox{ }m}^{-3}$ [$\rho_{\rm i}$ being calculated automatically via Equation (\ref{eq:densityrat})]. We select $k=5\times10^{-8}\mbox{ }{\rm m}^{-1}$ such that the incoming FMA wave has a period of about $100\mbox{ }{\rm s}$ (consistent with the order of magnitude of the period of observed EIT waves in the solar corona). It should be mentioned that the analysis presented in Figure \ref{fig:linearalfven} is valid for any $k$ as long as the dimensionless quantity $kx_0$ satisfies the condition $kx_0\ll1$.

Figure \ref{fig:linearalfven} gives the wave energy absorption coefficient for $kx_0=0.01$, $0.1$, $0.5$ and $1.0$ (the last one not being full consistent with our assumptions, but illustrates the trend). The angle of the incoming FMA wave, $\phi$, takes values between $0$ and $\pi/2$, while the angle of the equilibrium magnetic field, $\alpha$, only varies between $0$ and $\pi/4$, because beyond this point the integrals calculated are divergent and the numerical analysis cannot resolve the Cauchy principal part. The coefficient of wave energy absorption should take a value between $0$ and $1$, and can be thought of the percentage of incoming wave energy transferred to local \alf waves by resonance. We recognize that the plasma is unstable when the value of the coefficient of wave energy absorption is negative, provided a flow exists. The plasma can also create over-reflection, and we observe this in the coefficient of wave energy absorption when its value becomes greater than unity. Both the phenomena of over-reflection and instabilities occur for $\alpha>\pi/4$.

The first thing to notice, about Figure \ref{fig:linearalfven}, is that as $kx_0$ increases so does the coefficient of wave energy absorption. In the same way, the area (in the $\alpha$, $\phi$ plane) over which absorption can take place is also increasing. In general, we can state that the most efficient absorption occurs at angles of inclination of $\phi$ which are larger than $\pi/4$, while the most efficient absorption with respect to $\alpha$ changes depending on $kx_0$. The larger the value of $kx_0$ the less influential $\alpha$ becomes on the most efficient absorption regions. At approximately $\alpha=\pi/5\approx0.63$ the absorption drops to zero for all values of $kx_0$ and $\phi$, which could be explained by the inhomogeneous layer becoming \emph{transparent} to the incoming wave, a phenomenon that cannot be explained by the present model (further study is needed to find out whether it is a numerical artefact or a physical property).

It is also clear from Figure \ref{fig:linearalfven} that FMA waves are absorbed efficiently at the \alf resonance, which is encouraging when thinking about EIT waves within the solar corona. The EIT waves could be absorbed by \alf resonances present in / or near coronal loops (arcades). In some cases all of the energy of the incoming wave can be absorbed, and dissipated by, \emph{e.g.} viscosity. The variation in absorption due to combinations of $\alpha$ and $\phi$ may help explain why when an incoming wave impacts a coronal arcade, some loops oscillate more than others and some loops dim and while others get brighter. When the frequency of the incoming FMA wave does not fall within the frequencies of the \alf continuum the energy of the incident wave is, likely to be, transferred to the coronal loop as kinetic energy, thereby setting the loop into oscillation. These oscillations are studied in the framework of coronal seismology for the purposes of field and plasma diagnostics.

We note here that, if the density is varied (within reasonable parameters), the absorption rate is changed only slightly. If we change the density so it is really high or low (for the solar corona) the absorption starts dramatically changing, eventually leading to a breakdown of the numerical analysis. We do not show any plots of the variation of the absorption coefficient with density because the values at which the coefficient of wave energy absorption is noticeably changed are not consistent with observation of the solar corona (or chromosphere). The last significant variables to discuss are the equilibrium wave speeds. The characteristic wave speeds used here have been selected to be consistent with the environment of the solar corona. However, we know that there is a plethora of wave speeds available in the inhomogeneous plasma of the corona. We have tested different values of equilibrium speeds, and the overall pattern of absorption is identical to that found in Figure \ref{fig:linearalfven}, and the absorption coefficient remained relatively similar to those discussed above.

\subsection{Coupled Resonance: Modelling Chromospheric Absorption}

\begin{figure*}[!htb]
  \centering
  \includegraphics[width=5.5cm]{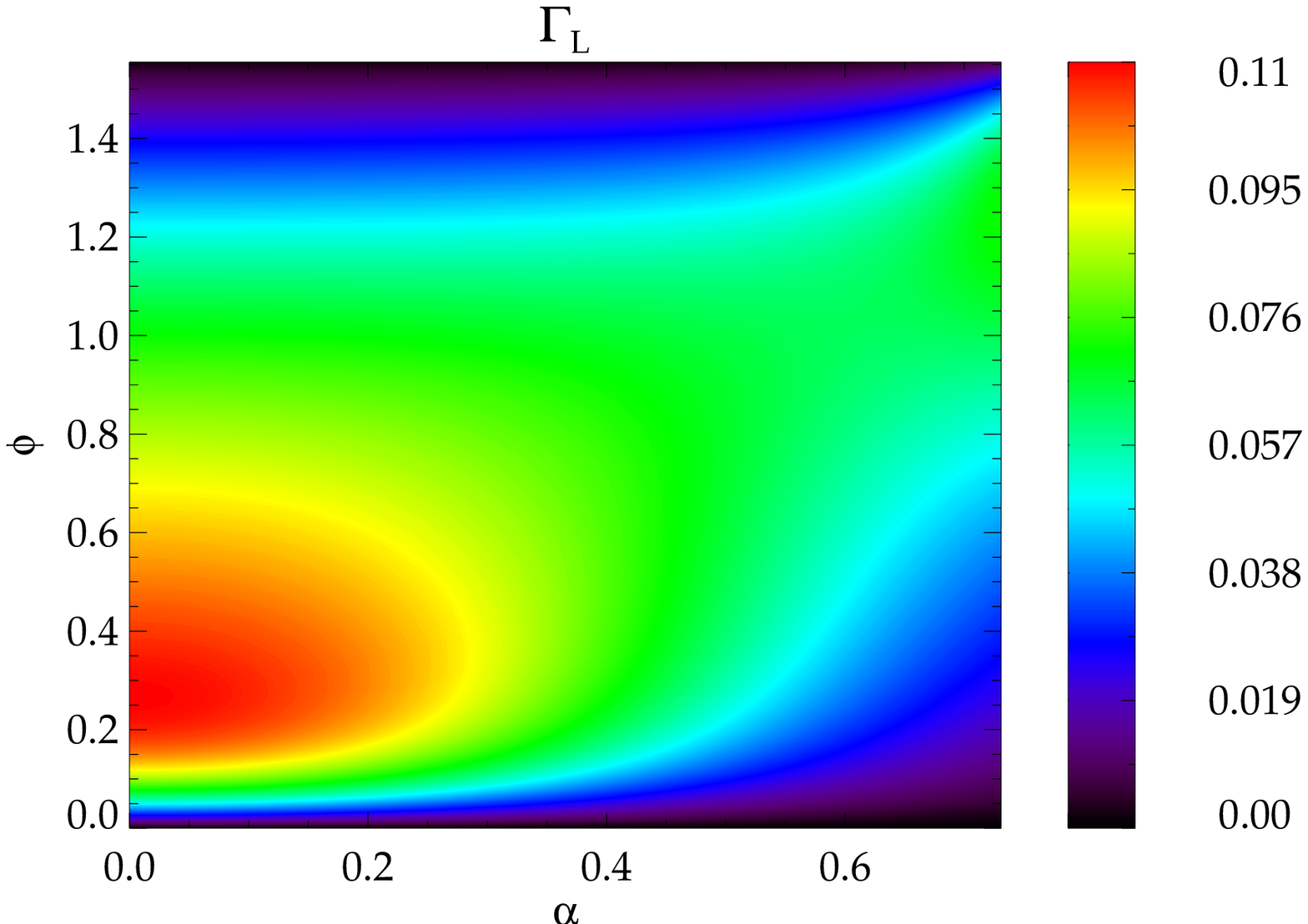}\hspace{2ex}\includegraphics[width=5.5cm]{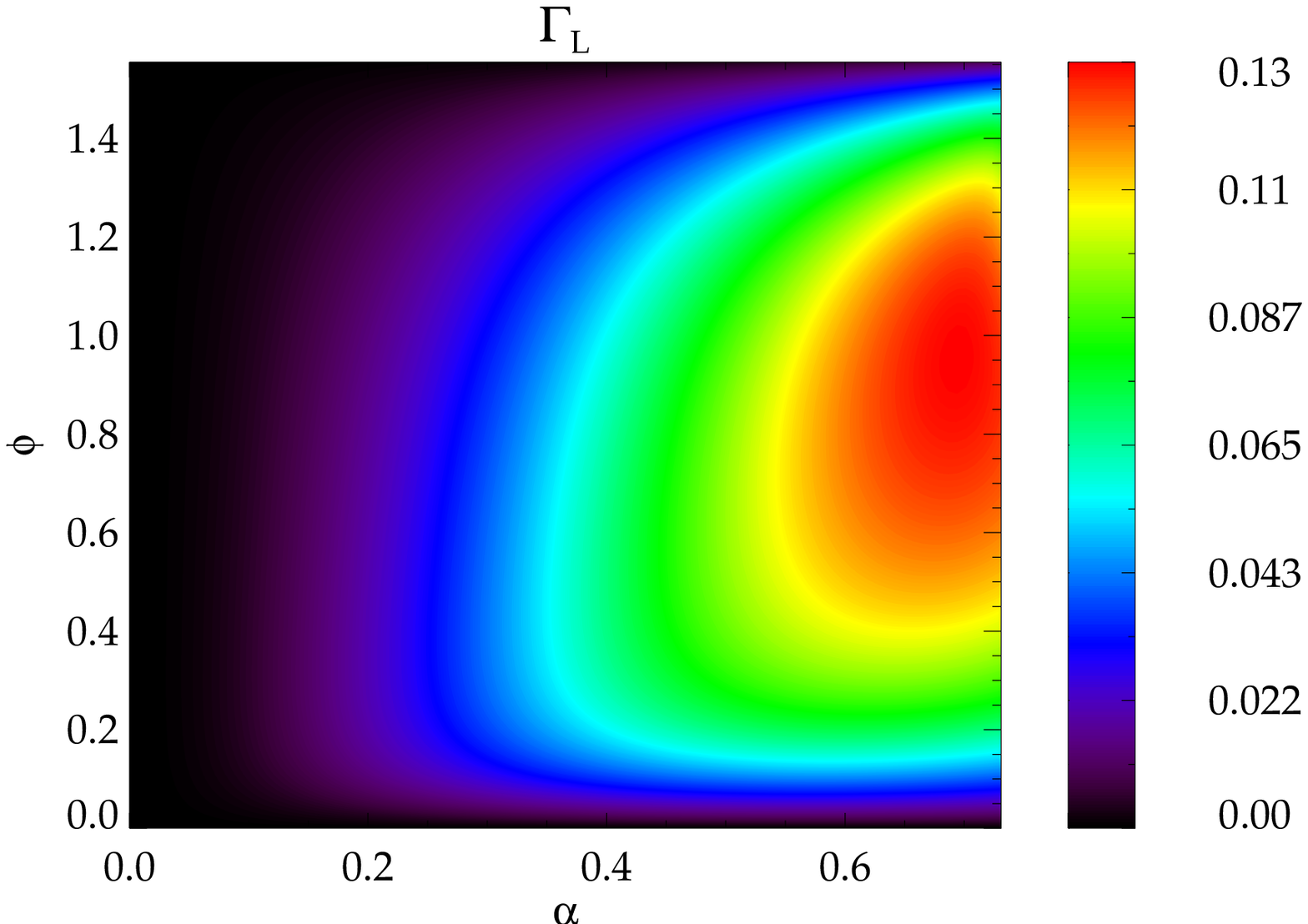}\\
  \includegraphics[width=5.5cm]{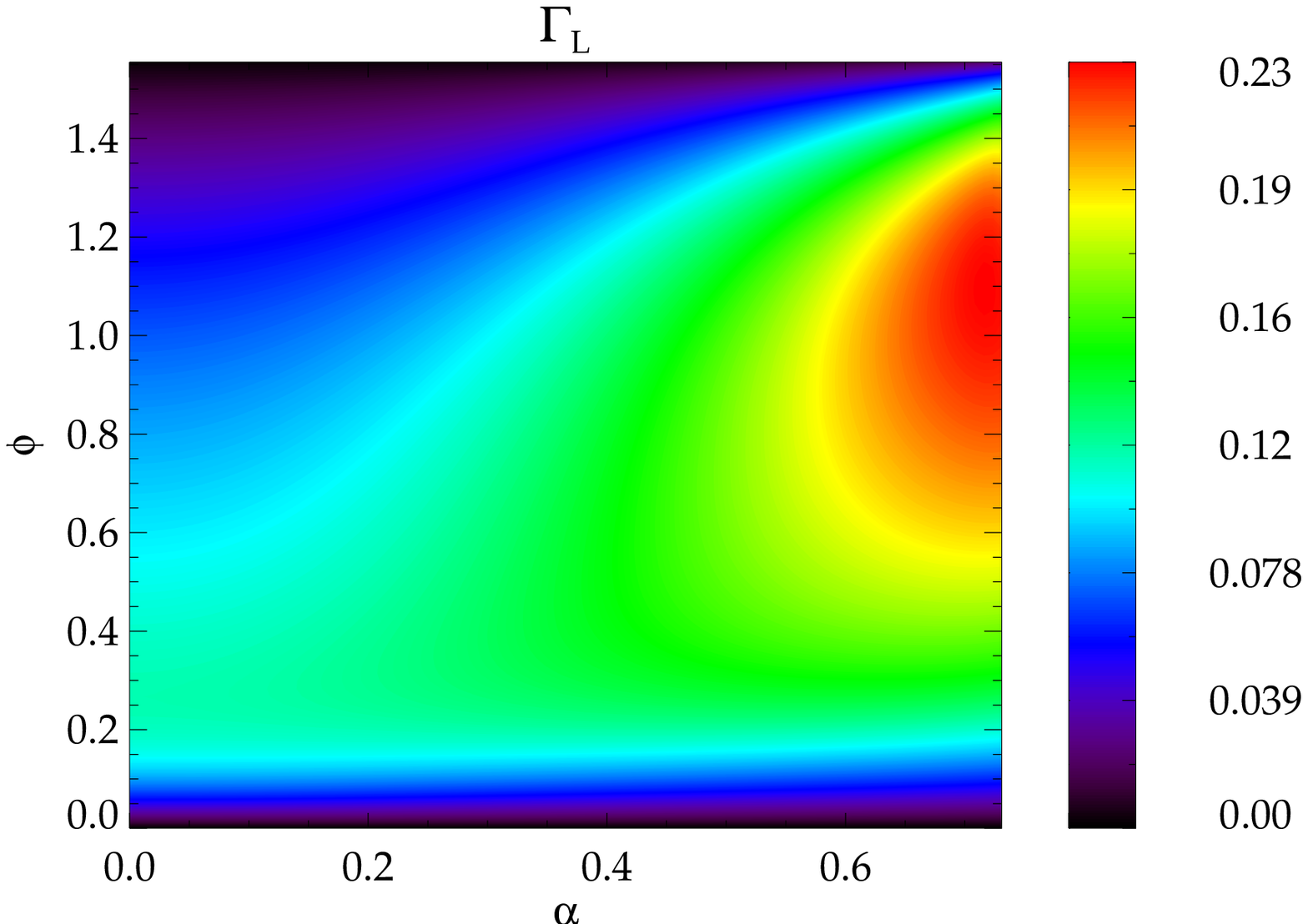}\hspace{2ex}\includegraphics[width=5.5cm]{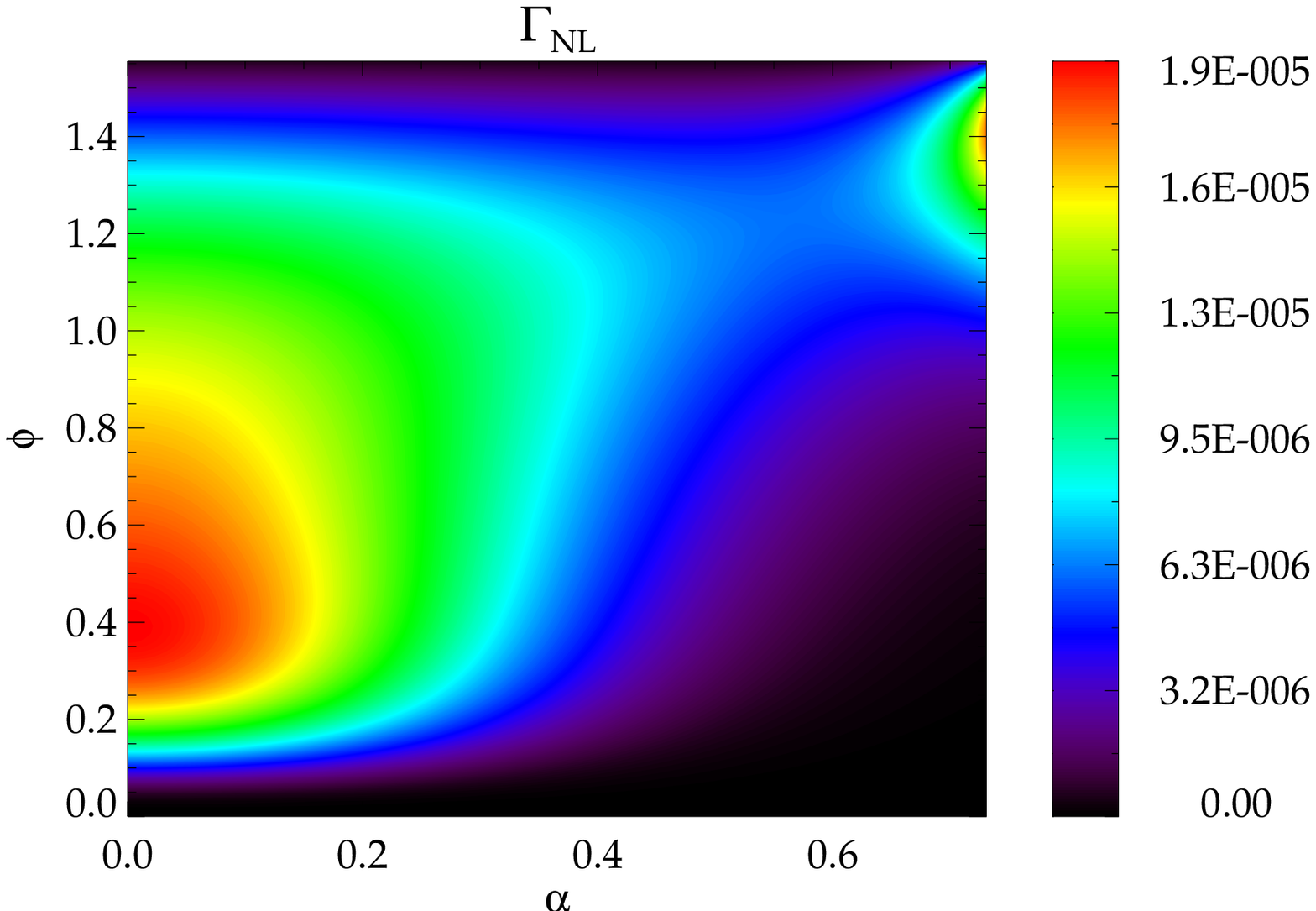}\\
  \caption{Comparison of the linear absorption at the slow (top left), \alf (top right) and coupled (bottom left) resonances for $kx_0=0.5$. We also show the nonlinear absorption for the coupled (and slow) resonance.} \label{fig:linearcouplekx0=0.5}
\end{figure*}

To model a coupled resonance, we consider a slow and an \alf resonance so close together that the local waves can both interact with the incoming FMA wave. This means that we will lower our applicability region to the denser chromosphere. To match conditions typical of the chromosphere we adopt the values $v_{\rm Ae}=28\mbox{ }{\rm km\mbox{ }s}^{-1}$, $c_{\rm Se}=34\mbox{ }{\rm km\mbox{ }s}^{-1}$, $v_{\rm Ai}=156\mbox{ }{\rm km\mbox{ }s}^{-1}$, $c_{\rm Si}=65\mbox{ }{\rm km\mbox{ }s}^{-1}$ and $\rho_{\rm e}=3.99\times10^{-11}\mbox{ }{\rm kg\mbox{ }m}^{-3}$. The specific values of equilibrium quantities can be changed, however, the overall variation trend of the coefficient of absorption will remain the same as shown here. The nonlinear correction to linear absorption at the coupled resonance comes solely from the slow resonance (as the \alf resonance can be described by linear theory) and is very small when compared to the linear absorption coefficient (in line with the weak nonlinear limit imposed in the derivations). In Figure \ref{fig:linearcouplekx0=0.5} we display the linear coefficient of wave energy absorption for the slow (top left panel), \alf (top right panel) and coupled resonance (bottom left panel) for comparison together with the nonlinear correction to the coupled resonance (bottom right panel) for values of $kx_0=0.5$. We have chosen to only display one value of $kx_0$ because the plots for other values of $kx_0$ look similar to this one, the only difference is the magnitude of the coefficient of wave energy absorption where, in general, the greater $kx_0$ the higher rate of absorption.

We have truncated the abscissa at $\alpha=6\pi/25$, because the integrals become divergent and the numerical analysis cannot resolve the Cauchy principal part. In Figure \ref{fig:linearcouplekx0=0.5} the nonlinear coefficient of wave energy absorption is incredibly small since the ratio of magnetic and plasma pressures (the plasma-$\beta$) has a value smaller than unity. When $\beta\ll1$ the plasma pressure is low and, by extension, [examine Equation (\ref{eq:coeffmajor})] so is the nonlinear coefficient of wave energy absorption. In addition, the nonlinear coefficient must be multiplied by the very small parameter $\zeta^2$. Therefore, the nonlinear correction of wave energy absorption at coupled resonances in the chromosphere is truly tiny in comparison to the linear wave energy absorption and acts to decrease the total coefficient of wave energy absorption.

In general, we can state that the greatest linear wave energy absorption at the slow resonance occurs at small values of $\phi$ ($0<\phi<0.3$) and small to moderate values of $\alpha$ ($0.1<\alpha<0.6$), whereas the the greatest linear wave energy absorption at the \alf resonance occurs at moderate values of $\alpha$ ($0.6<\alpha<0.8$) and a wide range of values of $\phi$ ($0.2<\phi<1.2$). For all values of $kx_0$ a larger percentage of energy is absorbed at the \alf resonance compared to the slow resonance, which is indicated by the coefficient of wave energy absorption having greater values at the \alf resonance. We can also state that increasing $kx_0$ produces more absorption (of course up to the point where the mathematical model becomes invalid).

The most interesting feature of Figure \ref{fig:linearcouplekx0=0.5} is the coupled resonance. When there is no absorption at both the slow and \alf resonance, the absorption at the coupled resonance is also zero and if there is absorption at only one of the slow or \alf resonance the coefficient of wave energy absorption at the coupled resonance is identical to the value at the single resonance. However, more interestingly, when there is absorption at both the single \alf and slow resonance the absorption at the coupled resonance is always \emph{greater} than the sum of the two single absorption coefficients. The greater coefficient of wave energy absorption at the coupled resonance compared to the two single resonances implies that there is more energy available for heating the plasma. We shall use Figure \ref{fig:linearcouplekx0=0.5} to clarify what we have just explained. At the values of, \emph{e.g.} $\alpha=0,\phi=\pi/2$ in all three plots we can clearly see the coefficient of wave energy absorption is zero. When, \emph{e.g.} $\alpha=0,\phi=0.4$ the coefficient of wave energy absorption at the \alf resonance is zero, while at the slow and coupled resonance it is $0.11$. If we change the angles such that, \emph{e.g.} $\alpha=6\pi/25,\phi=1.0$ the coefficient of wave energy absorption at the slow, \alf and coupled resonances are $0.067$, $0.13$ and $0.23$, respectively, hence it is clear that $0.23>0.13+0.067=0.197$. The same procedure can be carried out at any values of $\alpha$ and $\phi$ where there is absorption at both the single \alf and slow resonances.

There is also clear evidence in these plots that the efficiency at which FMA waves are absorbed at the \alf resonance under chromospheric conditions is far lower that in the coronal counterpart.

\section{Conclusions}

In the present paper, we have numerically investigated the absorption of fast magnetoacoustic (FMA) waves at individual and coupled slow and Alfv\'{e}n resonances, using the theory developed by \cite{clack2009a} where the coefficients of wave energy absorption were derived analytically by applying the long wavelength approximation ($kx_0\ll 1$).

We have shown that the absorption of incoming FMA waves depends heavily on the combination of the angle of incidence
of the wave ($\phi$) and the angle of inclination of the equilibrium magnetic field ($\alpha$). At some combinations of $\phi$ and $\alpha$, the quantity $\kappa^2_{\rm i}<0$ which reduces absorption to zero, because the incident FMA wave can \emph{leak} passed the resonances. Normally, the inhomogeneous layer is translucent to waves, but occasionally when the conditions are right, the inhomogeneous layer can become transparent so that they can pass through without undergoing resonant absorption.

We introduced the concept of coupled resonances which could be important for the heating of the solar upper atmosphere, because FMA waves propagate throughout the solar atmosphere. When FMA waves interact with local waves at a slow resonance, an \alf resonance is \emph{always} present as well (as long as there is an angle between the magnetic field and the direction of wave propagation). We believe this is the most likely form of resonant absorption in the upper chromosphere. In the corona, the Alfv\'{e}n (and hence FMA) waves speeds are much larger then the slow speed, therefore, a laterally driven incoming FMA wave can encounter an \alf resonance only. The governing equations for the slow and \alf resonances are derived considering nonlinearity, and even though nonlinearity only slightly decreases the resonant absorption, in comparison to linear theory, it provides a device by which the heated plasma can be transported: the \emph{mean shear flow}, creating turbulence, which can distort the inhomogeneous layer (see, \emph{e.g.} \opencite{ofman1995,clack2009b}), enhancing absorption and transporting heated plasma away from the resonance.

We assumed, for simplicity, that the equilibrium quantities inside the inhomogeneous layer increased monotonically, which, in reality, is not always the case. If we allow the equilibrium quantities to vary non-monotonically, the situation is changed. If we still use the long wavelength approximation, the calculations are almost identical to the ones produced in the present paper, however, instead of one position for the Alfv\'{e}n and slow resonances (different for each one) there could be several. Again, we would not be able to separate the individual outgoing waves, but the collective outgoing wave would be defined exactly as Equation (\ref{eq:use1}), where $\tau$ would equal the addition of all the $\tau_{\rm a}$s and $\tau_{\rm c}$s associated with the different resonant points. The form of the $\tau_{\rm a}$s and $\tau_{\rm c}$s would be identical to Equations (\ref{eq:taumuupsilon}) and (\ref{eq:tauc}), respectively, though the quantities would take different values corresponding to the different resonant positions. In general, this would produce greater absorption of wave energy inside the inhomogeneous layer, creating further heating possibility.

Extreme ultraviolet (EUV) observations of active regions showed that, when coronal arcades oscillate under the influence of an external driver, some of the loops oscillate more than others, some of the loops do not oscillate at all and some become \emph{dimmer} in the wavelength that they have been observed in. Part of this behaviour can be explained by the varying strength of the magnetic field inside the loops, however, this is not fully satisfactory. We \emph{speculate} that some of this phenomenon can be explained by resonant absorption. The varying conditions allow for varying degrees of resonant absorption. The loops that oscillate the most do not have the conditions necessary for resonant absorption, so the incident waves just transfer their kinetic energy to the loops. Other loops have the right conditions for resonant absorption, and some even have the conditions for multiple resonant positions. These loops will oscillate less, because the wave energy is being absorbed. The loops with the conditions for multiple resonances will oscillate the least (possibly not at all) as more and more energy of the wave is deposited. These loops should either get \emph{brighter} or \emph{dimmer}, since the loops are only observed in a single wavelength. If a loop \emph{brightens}, the plasma is emitting more intensely in the filters wavelength, but if a loops \emph{dims} the loop could be getting either cooler or hotter, as the plasmas emissions move out of the filters range. This explanation covers all the observed properties of some coronal arcade oscillations, and could explain why they are damped so quickly. The speculation must be viewed with caution as the heating due to absorption is likely to take place over a small area, in comparison to the width of the loop. We also add, however, that much more investigation is needed to substantiate this idea.

\begin{acknowledgements}

CTMC would like to thank STFC (Science and Technology Facilities
Council) for the financial support provided.  IB was financially supported by NFS Hungary (OTKA, K67746) and The National University
Research Council Romania (CNCSIS-PN-II/531/2007). The authors would also like to thank M.~S. Ruderman for his helpful advice and many interesting discussions.
\end{acknowledgements}
%
%
\bibliographystyle{apalike}
\bibliography{thesisreferences}

\begin{thebibliography}{}

\bibitem[Aschwanden, 1999]{aschwanden1999c}
Aschwanden, M.~J. (1999).
\newblock {Do EUV nanoflares account for coronal heating?}
\newblock {\em Solar Phys.}, 199:233.

\bibitem[Aschwanden et~al., 1999]{aschwanden1999}
Aschwanden, M.~J., Fletcher, L., Schrijver, C.~J., and Alexander, D. (1999).
\newblock {Coronal loop oscillations observed with the transition region and
  coronal explorer}.
\newblock {\em Astrophys. J.}, 520:880.

\bibitem[Athay and White, 1978]{athay1978}
Athay, R.~G. and White, O.~R. (1978).
\newblock {Chromospheric and coronal heating by sound waves}.
\newblock {\em Astrophys. J}, 526:1026.

\bibitem[Ballai et~al., 2008]{ballai2008}
Ballai, I., Douglas, M., and Marcu, A. (2008).
\newblock {Forced oscillations of coronal loops driven by EIT waves}.
\newblock {\em Astron. Astrophys.}, 488:1125.

\bibitem[Ballai and Erd\'{e}lyi, 1998]{ballai1998a}
Ballai, I. and Erd\'{e}lyi, R. (1998).
\newblock {Resonant absorption of nonlinear slow MHD waves in isotropic steady
  plasmas - I. Theory}.
\newblock {\em Solar Phys.}, 180:65.

\bibitem[Ballai et~al., 2005]{ballai2005}
Ballai, I., Erd\'{e}lyi, R., and Pint\'{e}r, B. (2005).
\newblock {On the nature of coronal EIT waves}.
\newblock {\em Astrophys. J. Lett.}, 633:145.

\bibitem[Ballai et~al., 1998a]{ballai1998b}
Ballai, I., Erd\'{e}lyi, R., and Ruderman, M.~S. (1998a).
\newblock {Interaction of sound waves with slow dissipative layers in
  anisotropic plasmas in the approximation of weak nonlinearity}.
\newblock {\em Phys. Plasmas}, 5:2264.

\bibitem[Ballai et~al., 1998b]{ballai1998c}
Ballai, I., Ruderman, M.~S., and Erd\'{e}lyi, R. (1998b).
\newblock {Nonlinear theory of slow dissipative layers in anisotropic plasmas}.
\newblock {\em Phys. Plasmas}, 5:252.

\bibitem[Belien et~al., 1999]{belien1999}
Belien, A.~J.~C., Martens, P.~C.~H., and Keppens, R. (1999).
\newblock {Coronal heating by resonant absorption: the effects of chromospheric
  coupling}.
\newblock {\em Astrophys. J.}, 526:478.

\bibitem[Braginskii, 1965]{braginskii1965}
Braginskii, S.~I. (1965).
\newblock {Transport processes in a plasma}.
\newblock {\em Rev. Plasma Phys.}, 1:205.

\bibitem[Chen and Hasegawa, 1974]{Chen1974}
Chen, L. and Hasegawa, A. (1974).
\newblock {Plasma heating by spatial resonance of Alfv\'{e}n wave}.
\newblock {\em Phys. Fluids}, 17:1399.

\bibitem[Clack and Ballai, 2008]{clack2008}
Clack, C. T.~M. and Ballai, I. (2008).
\newblock {N}onlinear theory of resonant slow waves in anisotropic and
  dispersive plasmas.
\newblock {\em Phys. Plasmas}, 15:082310.

\bibitem[Clack and Ballai, 2009a]{clack2009b}
Clack, C. T.~M. and Ballai, I. (2009a).
\newblock {Mean shear flows generated by nonlinear resonant Alfv\'{e}n waves}.
\newblock {\em Phys. Plasmas}, 16:072115.

\bibitem[Clack and Ballai, 2009b]{clack2009a}
Clack, C. T.~M. and Ballai, I. (2009b).
\newblock {Nonlinear resonant absorption of fast magnetoacoustic waves in
  strongly anisotropic and dispersive plasmas}.
\newblock {\em Phys. Plasmas}, 16:0402305.

\bibitem[Clack et~al., 2009]{clack2009d}
Clack, C. T.~M., Ballai, I., and Ruderman, M.~S. (2009).
\newblock {On the validity of nonlinear Alfv\'{e}n resonance in space plasmas}.
\newblock {\em Astron. Astrophys.}, 494:317.

\bibitem[Cs\'{i}k et~al., 1998]{csik1998}
Cs\'{i}k, A., \u{C}ade\u{z}, V.~M., and Goossens, M. (1998).
\newblock {Effects of mass flow on resonant absorption and on over-reflection
  of magnetosonic waves in low-$\beta$ solar plasmas}.
\newblock {\em Astron. Astrophys.}, 339:215.

\bibitem[Erd\'{e}lyi and Ballai, 2001]{erdelyi2001}
Erd\'{e}lyi, R. and Ballai, I. (2001).
\newblock {Nonlinear resonant absorption of fast magnetoacoustic waves due to
  coupling into slow continua in the solar atmosphere}.
\newblock {\em Astron. Astrophys.}, 368:662.

\bibitem[Erd\'{e}lyi and Goossens, 1995]{erdelyi1995}
Erd\'{e}lyi, R. and Goossens, M. (1995).
\newblock {Resonant absorption of Alfv\'{e}n waves in coronal loops in
  visco-resistive MHD}.
\newblock {\em Astron. Astrophys.}, 294:575.

\bibitem[Erd\'{e}lyi and Goossens, 1996]{erdelyi1996A}
Erd\'{e}lyi, R. and Goossens, M. (1996).
\newblock {Effects of flow on resonant absorption of MHD waves in viscous MHD}.
\newblock {\em Astron. Astrophys.}, 313:664.

\bibitem[Goossens et~al., 2002]{goossens2002}
Goossens, M., Andries, J., and Aschwanden, M.~J. (2002).
\newblock {Coronal loop oscillations. An interpretation in terms of resonant
  absorption of quasi-mode kink oscillations}.
\newblock {\em {Astron. Astrophys.}}, 394:L39.

\bibitem[Goossens and Poedts, 1992]{goossens1992A}
Goossens, M. and Poedts, S. (1992).
\newblock {Linear resistive magnetohydrodynamic computations of resonant
  absorption of acoustic oscillations in sunspots}.
\newblock {\em Astrophys. J.}, 384:348.

\bibitem[Goossens et~al., 1995]{goossens1995}
Goossens, M., Ruderman, M.~S., and Hollweg, J.~V. (1995).
\newblock {Dissipative MHD solutions for resonant Alfv\'{e}n waves in
  1-dimensional magnetic flux tubes}.
\newblock {\em Solar Phys.}, 157:75.

\bibitem[Grossmann and Tataronis, 1973]{grossmann1973}
Grossmann, M. and Tataronis, J. (1973).
\newblock {Decay of MHD waves by phase mixing}.
\newblock {\em {Z. Phys.}}, {261}:217.

\bibitem[Hasegawa and Chen, 1976]{Hasegawa1976}
Hasegawa, A. and Chen, L. (1976).
\newblock {Kinetic processes in plasma heating by resonant mode conversion of
  Alfv\'{e}n wave}.
\newblock {\em {Phys. Fluids}}, {19}:1924.

\bibitem[Hollweg, 1985]{Hollweg1985}
Hollweg, J.~V. (1985).
\newblock {Viscosity in a magnetized plasma - Physical interpretation}.
\newblock {\em J. Geophys. Res.}, 90:7620.

\bibitem[Hollweg, 1988]{hollweg1988}
Hollweg, J.~V. (1988).
\newblock {Resonance absorption of solar p-modes by sunspots}.
\newblock {\em Astrophys. J.}, 335:1005.

\bibitem[Ionson, 1978]{ionson1978}
Ionson, J.~A. (1978).
\newblock {Resonant absorption of Alfv\'{e}nic surface waves and the heating of
  solar coronal loops}.
\newblock {\em Astrophys. J.}, 226:650.

\bibitem[Lou, 1990]{lou1990}
Lou, Y.-Q. (1990).
\newblock {Viscous magnetohydrodynamic modes and p-mode absorption by
  sunspots}.
\newblock {\em Astrophys. J.}, 350:452.

\bibitem[Mocanu et~al., 2008]{mocanu2008}
Mocanu, G., Marcu, A., Ballai, I., and Orza, B. (2008).
\newblock {The problem of phase mixed shear Alfv\'{e}n waves in the solar
  corona revisited}.
\newblock {\em Astron. Nachr.}, 329:780.

\bibitem[Nakariakov and Ofman, 2001]{nakariakov2001}
Nakariakov, V.~M. and Ofman, L. (2001).
\newblock {Determination of the coronal magnetic field by coronal loop
  oscillations}.
\newblock {\em Astron. Astrophys.}, 372:L53.

\bibitem[Nakariakov et~al., 1999]{nakariakov1999a}
Nakariakov, V.~M., Ofman, L., Deluca, E.~E., Roberts, B., and Davila, J.~M.
  (1999).
\newblock {TRACE observation of damped coronal loop oscillations: implications
  for coronal heating}.
\newblock {\em {S}cience}, 285:862.

\bibitem[Ofman and Davila, 1995]{ofman1995}
Ofman, L. and Davila, J.~M. (1995).
\newblock {Nonlinear resonant absorption of Alfv\'{e}n waves in three
  dimensions, scaling laws, and coronal heating}.
\newblock {\em J. Geophys. Res.}, 100:23427.

\bibitem[Parker, 1988]{parker1988}
Parker, E.~N. (1988).
\newblock {Nanoflares and the solar X-ray corona}.
\newblock {\em Astrophys. J.}, 330:474.

\bibitem[Poedts et~al., 1989]{poedts1989}
Poedts, S., Goossens, M., and Kerner, W. (1989).
\newblock {Numerical simulation of coronal heating by resonant absorption of
  Alfv\'{e}n waves}.
\newblock {\em Solar Phys.}, 123:83.

\bibitem[Poedts et~al., 1990a]{poedts1990a}
Poedts, S., Goossens, M., and Kerner, W. (1990a).
\newblock {On the efficiency of coronal loop heating by resonant absorption}.
\newblock {\em Astrophys. J.}, 360:279.

\bibitem[Poedts et~al., 1990b]{poedts1990b}
Poedts, S., Goossens, M., and Kerner, W. (1990b).
\newblock {Temporal evolution of resonant absorption in solar coronal loops}.
\newblock {\em Comput. Phys. Commun.}, 59:95.

\bibitem[Poedts et~al., 1990c]{poedts1990c}
Poedts, S., Kerner, W., and Goossens, M. (1990c).
\newblock {Numerical simulation of the stationary state of periodically driven
  coronal loops}.
\newblock {\em Comput. Phys. Commun.}, 59:75.

\bibitem[Priest, 1984]{priest1984}
Priest, E.~R. (1984).
\newblock {\em {S}olar {M}agnetohydrodynamics}.
\newblock Springer, Berlin.

\bibitem[Roberts et~al., 1984]{roberts1984}
Roberts, B., Edwin, P.~M., and Benz, A.~O. (1984).
\newblock On coronal oscillations.
\newblock {\em Astrophys. J.}, 279:857.

\bibitem[Roussev et~al., 2001a]{roussev2001a}
Roussev, I., Doyle, J.~G., Galsgaard, K., and Erd\'{e}lyi, R. (2001a).
\newblock {Modelling of solar explosive events in 2D environments. III -
  Observable consequences}.
\newblock {\em Astron. Astrophys.}, 380:719.

\bibitem[Roussev et~al., 2001b]{roussev2001b}
Roussev, I., Galsgaard, K., Erd\'{e}lyi, R., and Doyle, J.~G. (2001b).
\newblock {Modelling of explosive events in the solar transition region in a 2D
  environment. I - General reconnection jet dynamics}.
\newblock {\em Astron. Astrophys.}, 370:298.

\bibitem[Roussev et~al., 2001c]{roussev2001c}
Roussev, I., Galsgaard, K., Erd\'{e}lyi, R., and Doyle, J.~G. (2001c).
\newblock {Modelling of explosive events in the solar transition region in a 2D
  environment. II - Various MHD experiments}.
\newblock {\em Astron. Astrophys.}, 375:228.

\bibitem[Ruderman, 2000]{ruderman2000}
Ruderman, M.~S. (2000).
\newblock {Interaction of sound waves with an inhomogeneous magnetized plasma
  in a strongly nonlinear resonant slow-wave layer}.
\newblock {\em J. Plasma Phys.}, 63:43.

\bibitem[Ruderman et~al., 1997a]{ruderman1997i}
Ruderman, M.~S., Berghmans, D., Goossens, M., and Poedts, S. (1997a).
\newblock {Direct excitation of resonant torsional Alfven waves by footpoint
  motions}.
\newblock {\em Astron. Astrophys.}, 320:305.

\bibitem[Ruderman et~al., 1997b]{ruderman1997a}
Ruderman, M.~S., Goossens, M., Ballester, J.~L., and Oliver, R. (1997b).
\newblock {Resonant Alfv\'{e}n waves in coronal arcades driven by footpoint
  motions}.
\newblock {\em Astron. Astrophys.}, 328:361.

\bibitem[Ruderman et~al., 1997c]{ruderman1997b}
Ruderman, M.~S., Goossens, M., and Hollweg, J.~V. (1997c).
\newblock {Nonlinear theory of the interaction of sound waves with an
  inhomogeneous magnetized plasma in the resonant slow wave layer}.
\newblock {\em Phys. Plasmas}, 4:91.

\bibitem[Ruderman et~al., 1997d]{ruderman1997c}
Ruderman, M.~S., Hollweg, J.~V., and Goossens, M. (1997d).
\newblock {Nonlinear theory of resonant slow waves in dissipative layers}.
\newblock {\em Phys. Plasmas}, 4:75.

\bibitem[Ruderman and Roberts, 2002]{ruderman2002}
Ruderman, M.~S. and Roberts, B. (2002).
\newblock {The damping of coronal loop oscillations}.
\newblock {\em Astrophys. J.}, 577:475.

\bibitem[Sakurai et~al., 1991]{sakurai1991}
Sakurai, T., Goossens, M., and Hollweg, J.~V. (1991).
\newblock {Resonant behaviour of MHD waves on magnetic flux tubes. I -
  Connection formulae at the resonant surfaces}.
\newblock {\em Solar Phys.}, 133:227.

\bibitem[Spruit and Bogdan, 1992]{stenuit1995}
Spruit, H.~C. and Bogdan, T.~J. (1992).
\newblock {The conversion of p-modes to slow modes and the absorption of
  acoustic waves by sunspots}.
\newblock {\em Astrophys. J. Lett.}, 391:109.

\bibitem[Tataronis and Grossmann, 1973]{tataronis1973}
Tataronis, J. and Grossmann, M. (1973).
\newblock {Decay of MHD waves by phase mixing}.
\newblock {\em {Z. Phys.}}, {261}:203.

\bibitem[Terradas et~al., 2008]{terradas2008}
Terradas, J., Arregui, I., Oliver, R., Ballester, J.~L., Andries, J., and
  Goossens, M. (2008).
\newblock {Resonant absorption in complicated plasma configurations:
  applications to multistranded coronal loop oscillations}.
\newblock {\em {Astrophys. J.}}, {679}:1611.

\bibitem[Terradas et~al., 2010]{terradas2010}
Terradas, J., Goossens, M., and Ballai, I. (2010).
\newblock {The effect of longitudinal flow on resonantly damped kink
  oscillations}.
\newblock {\em {Astron. Astrophys.}}, {}:accepted.

\bibitem[\u{C}ade\u{z} et~al., 1997]{cadez1997}
\u{C}ade\u{z}, V.~M., Cs\'{i}k, A., Erd\'{e}lyi, R., and Goossens, M. (1997).
\newblock {Absorption of magnetosonic waves in presence of resonant slow waves
  in the solar atmosphere}.
\newblock {\em Astron. Astrophys.}, 326:1241.

\bibitem[Uchida et~al., 1973]{uchida1973}
Uchida, Y., Altschuler, M.~D., and Newkirk, G. (1973).
\newblock {Flare-Produced Coronal MHD-Fast-Mode Wavefronts and Moreton's Wave
  Phenomenon}.
\newblock {\em Solar Phys.}, 28:495.

\bibitem[Vasquez, 2005]{vasquez2005}
Vasquez, B.~J. (2005).
\newblock {Resonant absorption of an Alfv\'{e}n wave: hybrid simulations}.
\newblock {\em J. Geophys. Res.}, 110:A10S10.

\bibitem[Woodward and McKenzie, 1994a]{woodward1994a}
Woodward, T.~I. and McKenzie, J.~F. (1994a).
\newblock {Stationary MHD waves modified by Hall current coupling - I. Cold
  compressible flow}.
\newblock {\em Planet Space Sci.}, 42:463.

\bibitem[Woodward and McKenzie, 1994b]{woodward1994b}
Woodward, T.~I. and McKenzie, J.~F. (1994b).
\newblock {Stationary MHD waves modified by Hall current coupling - II.
  Incompressible flow}.
\newblock {\em Planet Space Sci.}, 42:481.

\end{thebibliography}
%
%
%
%

\end{article}
\end{document}